\documentclass[%
reprint,
amsmath,amssymb,
aps,
]{revtex4-2}
\usepackage{multirow}
\usepackage{comment}
\usepackage{color,soul}
\usepackage{bm}
\usepackage{lipsum}
\usepackage{mathrsfs}
\usepackage{hyperref}
\usepackage{graphicx}
\usepackage{makecell}

\newcommand{\stat}[1]
{\left|#1\right\rangle}

\newcommand{\bra}[1]
{\left\langle#1\right|}

\newcommand{\ip}[2]
{\langle{#1}\vert{#2}\rangle}

\newcommand{\abs}[1]
{\left\vert#1\right\vert}

\usepackage{graphicx}
\usepackage{dcolumn}
\usepackage{bm}

\begin{document}
	
	\preprint{APS/123-pED}
	
	\title{Steered discrete-time quantum walks for engineering of quantum states}

	\author{Gururaj Kadiri}
			\email{gururaj@igcar.gov.in}
	\affiliation{Materials Science Group, Indira Gandhi Centre for Atomic Research, Kalpakkam, Tamilnadu, 603102, India.}

	\begin{abstract}
We analyze the strengths and limitations of steered discrete time quantum walks in generating quantum states of bipartite quantum systems comprising of a qubit coupled to a qudit system. We demonstrate that not all quantum states in the composite space are accessible through quantum walks, even under the most generalized definition of a quantum step, leading to a bifurcation of the composite Hilbert space into ``walk-accessible states" and the ``walk-inaccessible states". We give an algorithm for generating any walk-accessible state from a simple-to-realize product state, in a minimal number of walk steps, all of unit step size. We further give a prescription towards constructing minimal quantum walks between any pair of such walk-accessible states. Linear optics has been a popular physical system for implementing coin-based quantum walks, where the composite space is built up of spin and orbital angular momenta of light beams. We establish that in such an implementation, all normalized quantum states are ``walk-accessible". Furthermore, any generalized quantum step can be implemented upto a global phase using a single q-plate and a pair of homogeneous waveplates. We then give a quantum walk based scheme for realizing arbitrary vector beams, using only q-plates and waveplates.
		
	\end{abstract}
	
	\maketitle

	\section{Introduction to quantum walks}
	The theory of classical random walks has been an indispensable tool to understand statistical phenomena, and has been ubiquitous in modeling of various stochastic processes\cite{kac1947random,berg2018random,codling2008random,scalas2006application,ibe2013elements}. In one-dimensional random walks, the walker tosses a coin and moves forward or backward by one unit, depending upon the outcome of the toss. 
	The quantum mechanical counterpart of this are the discrete time quantum walks  (DTQWs), where the walker and the coin obey the principles of quantum mechanics, and the motion of the walker and the toss of the coin are implemented as unitary transformations. Owing to quantum superposition effects, these walks display distinct statistical properties compared to their classical counterparts.
	Quantum walks have proved to be of immense utility in varied domains, like in quantum computation\cite{PhysRevA.48.1687,singh2021universal,PhysRevLett.102.180501}, quantum search \cite{portugal2013quantum}, quantum algorithms \cite{kendon2006random,campos2021quantum}, generating random numbers \cite{sarkar2019multi}, modeling topological phenomena  \cite{panahiyan2021toward,panahiyan2020controllable,cardano2016statistical,cardano2017detection,barkhofen2017measuring} etc.
	In the DTQWs, the walker is modeled as a qudit, belonging to a potentially infinite-dimensional space called the ``walk space'', while the coin is modeled as a qubit, belonging to  two-dimensional space called the ``coin-space''. In this context, another application of such quantum walks explored in recent years has been quantum state engineering (QSE) of high-dimensional quantum systems, which refers to constructing a desired quantum state starting from some simple-to-realize initial state. QSE is realized using DTQWs by carefully steering the quantum walk towards a certain state in the composite space and then projecting out the coin, such that the walker collapses into the desired state on the walk-space \cite{innocenti2017quantum,giordani2019experimental,suprano2021dynamical,suprano2021real,zhang2022arbitrary}. Such high-dimensional states have myriad applications, since, in general  more can be accomplished in a qudit quantum system than in a qubit quantum system \cite{lanyon2009simplifying,gedik2015computational,cozzolino2019high}. For instance, qudits are shown to be more powerful than qubits in tests exploring the foundations of quantum theory \cite{kaszlikowski2000violations,rungta2001qudit,collins2002bell}, in quantum machine learning \cite{schuld2015introduction} etc.  In quantum cryptography, such states are shown to be more resilient to noise, and also offer a larger secret key rate \cite{sheridan2010security,islam2017provably,etcheverry2013quantum,bechmann2000quantum,durt2004security}.

Quantum walks have been implemented in various physical systems\cite{wang2013physical}: NMR systems \cite{ryan2005experimental,du2003experimental}, on ion traps \cite{xue2009quantum,matjeschk2012experimental}, in circuit  QED\cite{zhou2019protocol} in neutral atoms \cite{karski2009quantum}, to name a few.
In addition to these, photonic systems have also proved to be a very viable platform for implementing DTQWs, in that they have been demonstrated under various settings like, time-bin encoding\cite{boutari2016large}, interferometers \cite{harris2017quantum,aspuru2012photonic}, and spin angular momentum and orbital angular momentum (SAM-OAM) of light beams.  \cite{zhang2010implementation,goyal2013implementing,sephton2019versatile}.

In this paper, we examine the viability of DTQWs in engineering of the quantum states of the composite-space itself instead of those of walk-space alone. In SAM-OAM optical implementation of the quantum walks, these states physically correspond to what are called vector beams, which are light beams having spatially varying polarization distribution across their transverse planes. Such vector light beams have found numerous applications in varied fields\cite{dorn2003sharper,sick2000orientational,zhan2009cylindrical,woerdemann2013advanced,rosales2018review,yang2021optical}.  In this paper, we recast the above-mentioned ``QSE using DTQWs'' formulation as a means of generating desired vector vortex beams starting from standard Gaussian light beams. \\
The paper is organized as follows: In section \ref{sec:background}, we give  theoretical background of the DTQWs, and then present mathematical formulation of QSE in DTQWs. Section \ref{sec:Genarlized DTQWs} is devoted towards characterizing the quantum states obtained by quantum walks, and the quantum walks themselves. In section \ref{sec:Targeted_QWs}, we present a deterministic recursive algorithm for generating walk-accessible composite states, starting from some easy-to-generate product states. In section \ref{sec:Simplifying_QWs}, we introduce the notion of simplification of a quantum walk. We establish that given a quantum walk it is possible to construct another ``simplified" quantum walk that is completely equivalent to it, but consisting of quantum steps of identical step size. Given a pair of walk-accessible states, we also provide here method for obtaining a ``minimal quantum walk" that transforms one to the other. In Section \ref{sec:Illustrations}, we give a few numerical illustrations of the results presented in this paper. As a practical implementation of these ideas, in Section \ref{sec:Optical_Implementation}, we provide the scheme for generating desired vector beams using targeted quantum walks. In Section \ref{sec:Conclusion}, we conclude the paper by summarizing the results. 
	\section{Theoretical background \label{sec:background}}
	For carrying out a DTQW, we identity a quantum system with two different degrees of freedom (DOFs) : a qubit DOF called the ``coin-space'' spanning the Hilbert space $H_c$ and a qudit DOF called the ``walk space'' whose states belong to a Hilbert space $H_w$. 
	An arbitrary state $\stat{\boldsymbol{w}}\in H_c$ in the standard basis is given by
	\begin{equation}
	\stat{\boldsymbol{w}}=w_0\stat{\boldsymbol{0}}+w_1\stat{\boldsymbol{1}}
	\label{eq:qubit_definition}
	\end{equation}
	where $w_0$ and $w_1$ are complex numbers. We associate with every state a unique state $\stat{\boldsymbol{w}_\perp}$ which is orthogonal to $\stat{\boldsymbol{w}}$ defined in terms of components of $\stat{\boldsymbol{w}}$ in the standard basis as:
	\begin{equation}
	\stat{\boldsymbol{w}_\perp}=-\bar{w_1}\stat{\boldsymbol{0}}+\bar{w_0}\stat{\boldsymbol{1}}
	\label{eq:qubit_perp}
	\end{equation}
where the overbar indicates complex conjugation. Any state orthogonal to $\stat{\boldsymbol{s}}$ will differ $\stat{\boldsymbol{s}_\perp}$ only by an overall phase factor. It is to be noted that $\stat{(\boldsymbol{s}_\perp)_\perp}$, the orthogonal state of $\stat{\boldsymbol{s}_\perp}$, is not $\stat{\boldsymbol{s}}$ but -$\stat{\boldsymbol{s}}$. 
 We use the symbol $\stat{\boldsymbol{s};m}$ to represent the product state $\stat{\boldsymbol{s}}\otimes\stat{m}$, where $\stat{\boldsymbol{s}}$ is a unit vector in $H_c$, and $\stat{m}$ is the $m^{th}$ standard basis vector in $H_w$.  

An arbitrary pure state in the composite space $H_c\otimes H_w$ is given by \cite{ekert1995entangled,pathak2013elements}
\begin{equation}
\stat{\boldsymbol{P}}=\sum_{m=b}^{m=e}p_m\stat{\boldsymbol{p}_m;m} 
\label{eq:arb_composite_state}
\end{equation}
where $\stat{\boldsymbol{p}_m}$ are unit vectors in the coin-space $H_c$, and $p_m$ are non-negative real numbers. Given a composite state of the form Eq. (\ref{eq:arb_composite_state}), we define the integer interval $[b,e]$ as its ``position span'', by which we mean that the probability amplitudes at all positions greater than $e$ or less than $b$ are all zero.

\subsection{Coin-based quantum walks}
A single step of the quantum walk is a product of these two operators:
\begin{equation}
\hat{T}=\hat{S}\left(\hat{C}\otimes \hat{I}_w\right),
\label{eq:Standard_T}
\end{equation}
where $\hat{C}$ is an SU(2) operator called the ``coin toss operator'':
\begin{equation}
\hat{C}(\alpha,\beta,\gamma)=\begin{pmatrix}
e^{i\alpha}\cos\beta & -e^{-i\gamma}\sin\beta\\
e^{i\gamma}\sin\beta & e^{-i\alpha}\cos\beta\\
\end{pmatrix}
\label{eq:standard_W}
\end{equation}
in the $\{\stat{\boldsymbol{0}},\stat{\boldsymbol{1}}\}$ basis of the coin-space. 
The shift operator is given by:\begin{equation}
\hat{S}=\sum_{m}^{}\left(\stat{\boldsymbol{0};m-1}\bra{\boldsymbol{0};m}+\stat{\boldsymbol{1};m+1}\bra{\boldsymbol{1};m}\right)
\label{eq:standard_Shift_Operator}
\end{equation}
 
The initial state, that is, the state at the beginning of the quantum walk is generally taken to be at position $m=0$, in some arbitrary coin-state $\stat{\boldsymbol{s}}$. A quantum walk constitutes $N$ such steps starting from this initial state leads to a composite state of the form Eq. (\ref{eq:arb_composite_state}):
\begin{equation}
\hat{T}^N\stat{\boldsymbol{s};0}=\stat{\boldsymbol{W}}
\label{eq:Standard_Walk_definition}
\end{equation}
The state on the RHS $\stat{\boldsymbol{W}}$ is a composite state of the form Eq. (\ref{eq:arb_composite_state}). The role of coin operator leading to different composite states is studied in ref. \cite{chandrashekar2008optimizing}.

\subsection{Quantum State Engineering using quantum walks}
We refer to the product state located at the position $0$, that is states of the form $\stat{\boldsymbol{u};0}$, as the ``home-states". 
The problem of quantum state engineering is to generate a composite state, like Eq. (\ref{eq:arb_composite_state}), starting from some home-state. 
The state on the position space alone, after tracing out the coin degree of freedom from $\stat{\boldsymbol{S}}$ is a mixed state whose density matrix is given by 
\begin{equation}
	\rho=\textbf{tr}_{coin}(\stat{\boldsymbol{S}}\bra{\boldsymbol{S}})=\sum_{m=b}^{m=e}s_m^2\stat{\boldsymbol{m}}\bra{\boldsymbol{m}}
\end{equation}
In this manuscript, we intend to address some of the following questions: (i) can all quantum states on the composite space, that is, states of the form Eq. (\ref{eq:arb_composite_state}), be generated by a quantum walk of the form Eq. (\ref{eq:Standard_Walk_definition}) or some reasonable generalization of it, starting from some home-state? (ii) If two composite states can be generated by quantum walks, can those two be connected by a quantum walk? (iii) If the answer to the previous question is in affirmative, can a minimal walk be found that accomplishes it in least number of unit-sized steps?\\

\section{Generalized coin-based quantum walks
	 \label{sec:Genarlized DTQWs}}
 Speaking of generalizing the quantum walks, one method has been to employ different coin operator at each step, so that in place of Eq. (\ref{eq:Standard_Walk_definition}), we have:
 \begin{equation}
 \hat{T}_N\cdots\hat{T}_1\stat{\boldsymbol{s};0}=\stat{\boldsymbol{W}}
 \label{eq:Different_Coin_Walk_definition}
 \end{equation} 
 where $\hat{T}_i$ represents the $i^{th}$ quantum step, differing in their coin-flip operators, Eq. (\ref{eq:standard_W}). Another generalization of Eq. (\ref{eq:Standard_Walk_definition}) has been to have different step sizes for forward and backward motion. For instance, in refs. \cite{montero2013unidirectional,montero2015quantum,innocenti2017quantum,giordani2019experimental}, the conditional shift operator employed was the one of moving forward or staying still, depending upon the state of the coin. Owing to the translational symmetry of the problem, this shift operator is shown to be equivalent to the standard shift operator, in the sense that the resulting composite states in the both cases are identical, up to a relabeling of the position states. \\
 It is to be noted that replacing an SU(2) coin with a unitary one will not give access to any different set of states: a composite state accessible with a unitary coin is also accessible with an SU(2) coin. Therefore in the rest of the manuscript, we shall restrict our attention to only SU(2) coins. Further, there are generalizations of the discrete time quantum walks where the coins are site-dependent \cite{suzuki2016asymptotic,panahiyan2018controlling,zhang2022arbitrary} or history-dependent \cite{flitney2004quantum,rohde2013quantum,di2018elephant}. Such treatment are also beyond the scope of this manuscript, as we intend to consider only those systems which can be readily implemented in the linear optics setting. 

\subsection{Generalized quantum steps}
A single step of the quantum walk is an SU(2) operator $\hat{T}$ on the composite system $H_c\otimes H_w$ of the form: \begin{equation}
\hat{T}_\Gamma(\delta,p,\boldsymbol{s},\boldsymbol{c})=\hat{S}_\Gamma(\delta,p,\boldsymbol{s})\left(\hat{C}(\boldsymbol{s},\boldsymbol{c})\otimes I_w\right),
\end{equation} where $\hat{C}(\boldsymbol{s},\boldsymbol{c})$ and $\hat{S}_\Gamma(\delta,p,\boldsymbol{s})$ are given by:
\begin{widetext}
\begin{equation}
	\begin{aligned}
		\hat{C}(\boldsymbol{s},\boldsymbol{c})&=\stat{\boldsymbol{s}}\bra{\boldsymbol{c}}+\stat{\boldsymbol{s}_\perp}\bra{\boldsymbol{c}_\perp},\\
		\hat{S}_\Gamma(\delta,p,\boldsymbol{s})&=\cos\frac{\Gamma}{2}I_c\otimes I_w+\sin\frac{\Gamma}{2}\sum_{m}^{}\left(e^{i\delta}\stat{\boldsymbol{s}_\perp;m+p}\bra{\boldsymbol{s};m}-e^{-i\delta}\stat{\boldsymbol{s};m-p}\bra{\boldsymbol{s}_\perp;m}\right)
	\end{aligned}
\label{eq:step_definition_abstract}
\end{equation}
\end{widetext}
where $\Gamma$ and $\delta$ are real numbers between $0$ and $2\pi$, $p$ is an nonzero integer and $I_c$ and $I_w$ are identity operators in the coin and walk space respectively. The coin-operator $\hat{C}(\boldsymbol{s},\boldsymbol{c})$ is completely equivalent to the coin operator $\hat{C}(\alpha,\beta,\gamma)$ defined in Eq. (\ref{eq:standard_W}), only parameterized differently. Here we have parameterized it by two unit vectors $\boldsymbol{c}$ and $\boldsymbol{s}$ of the coin-space. The action of coin-toss operator  $\hat{C}(\boldsymbol{s},\boldsymbol{c})$ is to unitarily transform the orthogonal  pair of states $\stat{\boldsymbol{c}}$ and $\stat{\boldsymbol{c}_\perp}$ to another orthogonal pair of states $\stat{\boldsymbol{s}}$ and $\stat{\boldsymbol{s}_\perp}$, respectively.
The shift operator $\hat{S}_\Gamma(\delta,p,\boldsymbol{s})$ leaves a fraction of the state unaltered, and with the remaining fraction, it moves forward in the position space by $p$ units if the coin-state is $\stat{\boldsymbol{s}}$, and moves backwards by the same number of units if the coin-state is $\stat{\boldsymbol{s}_\perp}$.  Simultaneously, it also  transforms the coin-states $\stat{\boldsymbol{s}}$ and $\stat{\boldsymbol{s}_\perp}$ into $e^{i\delta}\stat{\boldsymbol{s}_\perp}$ and $-e^{-i\delta}\stat{\boldsymbol{s}}$ respectively. 
The action of  $\hat{T}_\Gamma(\delta,p,\boldsymbol{s},\boldsymbol{c})$ on the orthogonal states $\stat{\boldsymbol{c};m}$ and $\stat{\boldsymbol{c}_\perp;m}$ is given by:
\begin{equation}
	\begin{aligned}
		\hat{T}_\Gamma(\delta,p,\boldsymbol{s},\boldsymbol{c})\stat{\boldsymbol{c};m}&=\cos\frac{\Gamma}{2}\stat{\boldsymbol{s};m}+\sin\frac{\Gamma}{2}e^{i\delta}\stat{\boldsymbol{s}_\perp;m+p},\\
		\hat{T}_\Gamma(\delta,p,\boldsymbol{s},\boldsymbol{c})\stat{\boldsymbol{c}_\perp;m}&=\cos\frac{\Gamma}{2}\stat{\boldsymbol{s}_\perp;m}-\sin\frac{\Gamma}{2}e^{-i\delta}\stat{\boldsymbol{s};m-p}
	\end{aligned}
\label{eq:Step_definition}
\end{equation}
We shall regard Eq. (\ref{eq:Step_definition}) as the definition of the quantum step $\hat{T}_\Gamma(\delta,p,\boldsymbol{s},\boldsymbol{c})$, being completely equivalent to Eq. (\ref{eq:step_definition_abstract}). Its action of $\hat{T}_\Gamma(\delta,p,\boldsymbol{s},\boldsymbol{c})$ on any other product state $\stat{\boldsymbol{w};m}$ can obtained by expressing $\stat{\boldsymbol{w}}$ in the basis $(\stat{\boldsymbol{c}},\stat{\boldsymbol{c}_\perp})$ and acting it linearly on them. Its action on an arbitrary composite state can be found by acting it on each of the constituent terms. In the quantum step $\hat{T}_\Gamma(\delta,p,\boldsymbol{s},\boldsymbol{c})$, we call the state $\stat{\boldsymbol{c}}$ as its coin-state and $\stat{\boldsymbol{s}}$ as its shift state. Here, $p$ is an integer indicating the step size of the walk. The presence of $\Gamma$ makes the quantum step very different from that of Eq. (\ref{eq:Standard_T}) as only a fraction of the state now undergoes the quantum walk, with $\Gamma$ determining that fraction: $\Gamma=\pi$ leading to complete state transfer, while $\Gamma=0$ corresponds to no change in the walk space. Such walks have been termed ``hybrid walks" in literature \cite{cardano2015quantum}. While $\Gamma$ can range from $[0,4\pi)$, it is easy to see that $\hat{T}_\Gamma(\delta,p,\boldsymbol{s},\boldsymbol{c})=\hat{T}_{4\pi-\Gamma}(\pi+\delta,p,\boldsymbol{s},\boldsymbol{c})$ and therefore can be restricted to $\Gamma$ to $[0,2\pi]$. 
The quantum steps $\hat{T}_0(\cdot,\cdot,\boldsymbol{s},\boldsymbol{c})$, i.e., the steps with $\Gamma=0$ refer only to SU(2) transformation from $\boldsymbol{c}$ to $\boldsymbol{s}$, acting only on the coin-space:
\begin{equation}
	\hat{T}_0(\cdot,\cdot,\boldsymbol{s},\boldsymbol{c})=\hat{C}(\boldsymbol{s},\boldsymbol{c})\otimes I_w
	\label{eq:p_0_definition}
\end{equation}
where $\hat{C}(\boldsymbol{s},\boldsymbol{c})$ is the operator defined in the first of  Eq. (\ref{eq:step_definition_abstract}). 
The step-size $p$ and the relative phase $\delta$ in this case are immaterial, and hence they are suppressed in Eq. (\ref{eq:p_0_definition}). We call these steps as the ``improper steps'', as against the steps with $\Gamma>0$ and $p\neq0$, which we call as the ``proper steps''.
The quantum step Eq. (\ref{eq:Step_definition}) is such that, in one proper step at most only two other states at a fixed distance ($\pm p$) from the current state, are accessed. We shall regard the step defined in Eq. (\ref{eq:Step_definition}) as the most generalized quantum walk step. This claim requires some justification, since we still have the same step-size $p$ in both forward and backward directions. This definition of a quantum step is the most general, in the sense that any further relaxation in the definition will be inconsistent with the requirement of translational invariance. 
It is essential, for instance, that the coin-states at the $m^{th}$ and $(m\pm p)^{th}$ positions appearing in the right hand side of Eq. (\ref{eq:Step_definition}) be orthogonal to each other. Likewise, having distinct step sizes in the forward and backward directions is not consistent with translational invariance, unless $\Gamma$ is fixed to $\pi$. In this manuscript, we have made a choice to force identical step size in the forward and backward directions, retaining the facility to have a variable $\Gamma$ instead. This choice is motivated by the linear optics implementation, to be introduced in Section \ref{sec:Optical_Implementation}. 
It must be noted that the second relation in Eq. (\ref{eq:Step_definition}) is not independent, but follows directly from the first, by requiring that $\hat{T}$ act as an SU(2) operator in the composite space. 
\\
We represent a collection of $M$ such steps by the symbol $\hat{W}$
\begin{equation}
\hat{W}=\hat{T}_{\Gamma_M}(\delta_M,p_m,\boldsymbol{s}_m,\boldsymbol{c}_m)\cdots \hat{T}_{\Gamma_1}(\delta_1,p_1,\boldsymbol{s}_1,\boldsymbol{c}_1)
\label{eq:Walk_definition}
\end{equation}
We call these sequences of such steps $\hat{W}$ as the ``quantum walk".  The walk $\hat{W}$ acts on the states from the left. We shall consider the dimension of walk-space to be larger than $N=\textbf{max}(\vert b \vert,\vert e \vert)$  of all the involved composite states, and also larger than $\vert p_1 \vert +\vert p_2 \vert +\cdots \vert p_M \vert $ of all the involved steps. 
\subsection{Identifying states obtained by quantum walks}
A quantum walk of $M$-steps, starting from an arbitrary home-state $\stat{\boldsymbol{u};0}$, can be described as:
\begin{equation}
\hat{W}\stat{\boldsymbol{u},0}=\stat{\boldsymbol{U}}=\sum_{m=b}^{m=e}u_m\stat{\boldsymbol{u}_m;m}
\label{eq:Walk_State_definition}
\end{equation} 
Note that not all steps in $\hat{W}$ are of equal step size. The position span $[b,e]$ of $\stat{\boldsymbol{U}}$ depends on the details of the steps, and also on the initial coin-state $\stat{\boldsymbol{u}}$. All that can be said is that $\textbf{max}(\vert b \vert,\vert e \vert)\le (\vert p_1 \vert +\vert p_2 \vert +\cdots \vert p_M \vert )$, where $p_i$ is the step size of the $i^{th}$ step. 
The state on the RHS is a unit vector of the form Eq. (\ref{eq:arb_composite_state}). We now derive a condition that such states have to satisfy. 
From translation symmetry, if $\hat{W}\stat{\boldsymbol{u};0}=\stat{\boldsymbol{U}}$ then $\hat{W}\stat{\boldsymbol{u};d}=\stat{\boldsymbol{U}_{+d}}$, 
where $d$ is an integer and $\stat{\boldsymbol{U}_{+d}}$ is $\stat{\boldsymbol{U}}$ shifted on the walk forward en-mass by $d$ units:
\begin{equation}
	\stat{\boldsymbol{U}_{+d}}=\sum_{n=b}^{n=e}u_n\stat{\boldsymbol{u}_n;n+d} 
\label{eq:Displaced_Definition}
\end{equation}
Since $\stat{\boldsymbol{u};0}$ and $\stat{\boldsymbol{u};d}$ are orthogonal, and since the walk operator $\hat{W}$ is an SU(2) operator, we demand the resulting states $\stat{\boldsymbol{U}}$ and $\stat{\boldsymbol{U}_{+d}}$  to also be orthogonal, $\ip{\boldsymbol{U}}{\boldsymbol{U}_{+d}}=0$, for all $d\neq0$. The non-trivial $d$ being $d=1,\cdots,e-b$, leading to $e-b$ constraints:
\begin{equation}
	\ip{\boldsymbol{U}}{\boldsymbol{U}_d}=\sum_{m=b}^{m=e-d} u_m u_{m+d}\ip{\boldsymbol{u_m}}{\boldsymbol{u_{m+d}}}=0,  \forall\:d=1,\cdots,e-b
\label{eq:TI_Constraints}
\end{equation}
This result limits the composite states that can be obtained by quantum walks: not all states of the composite space are accessible by quantum walks. We shall refer to the composite states satisfying the constraints of Eq. (\ref{eq:TI_Constraints}) as the ``walk-states'', and those composite states not satisfying them as the ``non-walk-states''.
For instance, consider the product states in the composite space $H$. They are of the form $\stat{\boldsymbol{u}}\otimes\stat{\boldsymbol{s}}$, where $\stat{\boldsymbol{u}}\in \textbf{span}(\stat{\boldsymbol{0}},\stat{\boldsymbol{1}})$ and $\stat{\boldsymbol{s}}\in \textbf{span}(\cdots,\stat{\boldsymbol{-1}},\stat{\boldsymbol{0}},\stat{\boldsymbol{1}},\cdots)$. Of these, the only possible translation invariant product states are of the form $\stat{\boldsymbol{u};m}$, where $\stat{m}\in (\cdots,\stat{\boldsymbol{-1}},\stat{\boldsymbol{0}},\stat{\boldsymbol{1}},\cdots)$. 
Likewise, a composite state with only two non-zero OAM components, $\stat{\boldsymbol{W}}=w_b\stat{\boldsymbol{w}_b;b}+w_e\stat{\boldsymbol{w}_e;e}$ is a walk-state only if the two constituent coin-states are orthogonal: $\ip{\boldsymbol{w}_b}{\boldsymbol{w_e}}=0$.
Likewise, a composite state $\stat{\boldsymbol{X}}$  consisting of three constituent terms: $\stat{\boldsymbol{X}}=x_b\stat{\boldsymbol{x}_b;b}+x_m\stat{\boldsymbol{x}_m;m}+x_e\stat{\boldsymbol{x}_e;e}$
where $b$, $m$ and $e$ are integers and $x_b$, $x_m$ and $x_e$ are real numbers such that $x_b^2+x_m^2+x_e^2=1$, is a walk-state only if
\begin{equation*}
	\begin{aligned}
		m&=\frac{b+e}{2},\\
		\stat{\boldsymbol{x}_m}&=\frac{1}{\sqrt{x_e^2+x_b^2}}\left(e^{i\delta}x_e\stat{\boldsymbol{x}_b}-e^{-i\delta}x_b\stat{\boldsymbol{x}_e}\right),\\
		\ip{\boldsymbol{x}_b}{\boldsymbol{x_e}}&=0
	\end{aligned}
\end{equation*} where $\delta$ is an arbitrary phase. This is an interesting result, with respect to the state engineering aspect of the quantum walks. The first of these restricts the possible occupied positions of a three-term walk-state. It is not possible via quantum walks to access composite states where, for instance, only the $\stat{\boldsymbol{0}}$, $\stat{\boldsymbol{1}}$ and $\stat{\boldsymbol{3}}$ position states are occupied. And this is true irrespective of the corresponding amplitudes of occupation $x_0$,  $x_1$ and $x_2$ and the coin-states  $\stat{\boldsymbol{x}_0}$, $\stat{\boldsymbol{x}_1}$ and $\stat{\boldsymbol{x}_3}$. Consider now the four component state \begin{equation*}
	\stat{\boldsymbol{Y}}=y_b\stat{\boldsymbol{y}_b;b}+y_m\stat{\boldsymbol{y}_m;m}+y_n\stat{\boldsymbol{y}_n;n}+y_e\stat{\boldsymbol{y}_e;e},
	\label{eq:four_composite_state}
\end{equation*}
where $b$, $m$, $n$ and $e$ are integers and $y_b$, $y_m$, $y_n$ and $y_e$ are non-zero real numbers such that $y_b^2+y_m^2+y_n^2+y_e^2=1$. It follows from Eq. (\ref{eq:TI_Constraints}) that $\stat{\boldsymbol{Y}}$ is a walk-state, if and only if $m-b=e-n$. Further, if $e-b\neq m-b(=e-n)$, then it is a walk-state only if the amplitudes satisfy the relation $\frac{y_b}{y_m}=\frac{y_e}{y_n}$,  and the coin-states satisfy the relations  $\ip{\boldsymbol{y}_b}{\boldsymbol{y}_m}=\ip{\boldsymbol{y}_m}{\boldsymbol{y}_n}=\ip{\boldsymbol{y}_n}{\boldsymbol{y}_e}=0$ and $\ip{\boldsymbol{y}_b}{\boldsymbol{y}_n}+\ip{\boldsymbol{y}_m}{\boldsymbol{y}_e}=0$. A few illustrations of walk and non-walk states will be given in sub-section (\ref{subsec:walk_non_walk}).
\subsection{Generating new walk-states from existing ones}
It must be noted that a linear combination of two walk-states need not be a walk-state. The set of walk-states, therefore do not form a subspace. This can be easily seen by noting that all the states belonging to the standard product basis $\{\stat{\boldsymbol{0}},\stat{\boldsymbol{1}}\}\otimes\{\cdots,\stat{\boldsymbol{-1}},\stat{\boldsymbol{0}},\stat{\boldsymbol{1}},\cdots\}$ are walk-states, and therefore if the linear combination of walk-states also yielded a walk-state, then every state in the composite space would have been a walk-state.\\
Given a walk-state $\stat{\boldsymbol{U}}$ with position span $[b,e]$, one can generate another walk-state $\stat{\boldsymbol{U}_{+d}}$ with span $[b+d,e+d]$ by shifting each position $m$ in $\stat{\boldsymbol{U}}$ by $d$ units. Likewise, we define another walk-state $\stat{\boldsymbol{U}_{\times d}}$ as:
\begin{equation}
\stat{\boldsymbol{U}_{\times d}}=\sum_{n=b}^{n=e}u_n\stat{\boldsymbol{u}_n;nd} 
\label{eq:U_times_d}
\end{equation}
It is easy to confirm that $\stat{\boldsymbol{U}_{\times d}}$ is also a walk-state. If a walk $\hat{W}$ generates $\stat{\boldsymbol{U}}$, then $\stat{\boldsymbol{U}_{\times d}}$ can be generated by changing the step size $p_i$ of each step  $\hat{T}_{\Gamma_i}(\delta_i,p_i,\boldsymbol{u}_i,\boldsymbol{c}_i)$ of $\hat{W}$ by $p_i\times d$.

Recall that the action of a linear operator on an $N$-dimensional space gets fixed only by specifying its action on $N$ linearly independent vectors. In case of the walk operator, however, once the action of $\hat{W}$ on any state $\stat{\boldsymbol{u};0}$ is specified, as in Eq. (\ref{eq:Walk_State_definition}), then its action on any other state, be it a home-state, a product-state or a composite-state, gets fixed, owing to its being SU(2) and translationally invariant. 
Particularly, between orthogonal home-states we have the following relation:
\begin{equation}
\begin{aligned}
\text{if } &	\hat{W}\stat{\boldsymbol{u};0}=\stat{\boldsymbol{U}}, \\\text{ then } &
\hat{W}\stat{\boldsymbol{u_\perp};0}=\stat{\boldsymbol{U}_\perp}
\end{aligned}
\label{eq:Ortho_ortho_mapping}
\end{equation}
where, for a given $\stat{\boldsymbol{U}}$, the composite state $\stat{\boldsymbol{U}_\perp}$ is defined as 
\begin{equation}
\stat{\boldsymbol{U}_\perp}=\sum_{m=-e}^{m=-b}u_{-m}\stat{(\boldsymbol{u}_{-m})_\perp;m}
\label{eq:U_perp}
\end{equation}
where $\stat{(\boldsymbol{u}_{-m})_\perp}$ is the state orthogonal to $\stat{\boldsymbol{u}_{-m}}$.   It is easy to see that the walk-states $\stat{\boldsymbol{U}}$ and $\stat{\boldsymbol{U}_\perp}$ are orthogonal to each other. Further, they share the same characteristic vector,  $\overrightarrow{p}(\boldsymbol{U})=\overrightarrow{p}(\boldsymbol{U}_\perp)$. 
Given a walk-state $\stat{\boldsymbol{U}}$, we can construct another walk-state $\stat{\boldsymbol{U}_\perp}$, and two other families of walk-states $\stat{\boldsymbol{U}_{+d}}$ and $\stat{\boldsymbol{U}_{\times d}}$ by shifting and scaling the walk position labels respectively. Furthermore, any complex linear combination of the walk-states $\stat{\boldsymbol{U}}$ and $\stat{{({\boldsymbol{U}}_\perp)}_{+d}}$ is also a legitimate walk-state for any $d$. This facility enables us to construct new walk-states of larger span from smaller ones. 
\subsection{Characterizing the walk-states}
Given two walk-states, we are interested in knowing whether they are a result of same walk $\hat{W}$, but with different home-states or not.   Let $\stat{\boldsymbol{P}}$ and $\stat{\boldsymbol{Q}}$ be the two walk-states originating from  home with coin-states $\stat{\boldsymbol{p}}$ and $\stat{\boldsymbol{q}}$ respectively, through a walk $\hat{W}$, that is, $\stat{\boldsymbol{P}}=\hat{W}\stat{\boldsymbol{p};0}$ and $\stat{\boldsymbol{Q}}=\hat{W}\stat{\boldsymbol{q};0}$. We now discuss the conditions that these walks states must satisfy if they are so related. The position-spans of $\stat{\boldsymbol{P}}$ and $\stat{\boldsymbol{Q}}$ are not of much help here: two composite states may well share the same position span, but might not have originated from the same walk, or they might have different position-spans but have originated from the same walk.  For instance, the first and second states on the RHS of Eq. (\ref{eq:Step_definition}) have different position spans,  $[b,b+m]$ and $[b-m,b]$ respectively, but they have originated from the same walk,  $\hat{T}$. One necessary condition that can be derived on the basis of position-span is that $N=\textbf{max}(\vert b \vert,\vert e \vert)$ of both the states be identical. 
We now derive another necessary condition. For this, we first recall that the states $\stat{\boldsymbol{p};0}$ and $\stat{\boldsymbol{q};d}$ are orthogonal in the composite state, for any $d\neq 0$, independent of the relation between the coin-states $\stat{\boldsymbol{p}}$ and $\stat{\boldsymbol{q}}$. Since $\hat{W}$ is an SU(2) operator, it preserves this orthogonality, leading to the constraints $	\ip{\boldsymbol{P}}{\boldsymbol{Q}_d}=0, \forall d \neq 0$, where $\stat{\boldsymbol{Q}_d}$ is $\stat{\boldsymbol{Q}}$ shifted by $d$ steps, as in Eq. (\ref{eq:Displaced_Definition}). 

For obtaining another necessary condition, given a composite state $\stat{\boldsymbol{U}}$, with the span $[b,e]$, we an define an $N+1$-dimensional probability mass function $\overrightarrow{p}(\boldsymbol{U})$, where $N=\textbf{max}(\vert b \vert,\vert e \vert)$ as:
\begin{equation}
	\overrightarrow{p}(\boldsymbol{U})=(f_0,f_1,f_2,\cdots,f_N)
	\label{eq:characteristic_function}
\end{equation}
where $f_m={u_m}^2+u_{-m}^2$, where $u_m$ are amplitudes at the position $m$ in the composite state $\stat{\boldsymbol{U}}$. From this definition, it is easy to verify that 
\begin{equation}
	\begin{aligned}
		\text{if } \stat{\boldsymbol{P}}=\hat{W}\stat{\boldsymbol{p};0} \text{ and }  \stat{\boldsymbol{Q}}=\hat{W}\stat{\boldsymbol{q};0} \text{ then } 
		\overrightarrow{p}(\boldsymbol{P})=\overrightarrow{p}(\boldsymbol{Q})
	\end{aligned}
	\label{eq:char_function_relation}
\end{equation} 
The three conditions offered above are only necessary for two walk-states to have emerged from same quantum walk, not sufficient. A necessary and also sufficient condition for two walk-states $\stat{\boldsymbol{P}}$ and $\stat{\boldsymbol{Q}}$ being output of the same walk is that $\abs{\ip{\boldsymbol{P}}{\boldsymbol{Q}}}^2+\abs{\ip{\boldsymbol{P}}{\boldsymbol{Q}_\perp}}^2=1$. \\
\subsection{Characterizing the quantum walks}
Having defined a ``characteristic vector" for a walk-state, we could, likewise, define it for a quantum walk. Given a  $\hat{W}$, we define its characteristic vector $\overrightarrow{p}(\hat{W})$ as 
\begin{equation}
	\overrightarrow{p}(\hat{W})=\overrightarrow{p}(\hat{W}\stat{\boldsymbol{u};0}) 
\end{equation}
for some arbitrary coin-state $\stat{\boldsymbol{u}}$. Observe that the $\overrightarrow{p}$ appearing on the LHS is that of a walk, while the one appearing on the RHS is that of a walk-state, defined in Eq. (\ref{eq:characteristic_function}). That this quantity is identical for any $\stat{\boldsymbol{u}}$ follows from Eq. (\ref{eq:char_function_relation}). The import of  characteristic vector of a walk is that it gives the description of its action in terms of walk space alone, without regard to the coin-space. Given an arbitrary product state $\stat{\boldsymbol{u};m}$ as input,  the walk $\hat{W}$ leaves a fraction $f_0$ of it unaltered, alters the position of a fraction $f_1$ of it by one unit, alters the position of another fraction $f_2$ of it by two units and so on. 
All $\Gamma=0$ steps have the characteristic vector $(1)$ and those with non-zero $\Gamma$ has the characteristic vector of length $p+1$:  $(\cos^2\frac{\Gamma}{2},\underbrace{0,\cdots,0}_{p-1},\sin^2\frac{\Gamma}{2})$. 
\section{Targeted quantum walks \label{sec:Targeted_QWs}}
It is to be noted that a quantum walk $\hat{W}$ preserves the translational invariance of composite states: it takes walk-states to walk-states and non-walk-states to non-walk-states.
Given a step $\hat{T}_\Gamma(\delta,p,\boldsymbol{s},\boldsymbol{c})$, we define its inverse $[\hat{T}_\Gamma(\delta,p,\boldsymbol{s},\boldsymbol{c})]^{-1}$ as the operator that undoes the action of $\hat{T}_\Gamma(\delta,p,\boldsymbol{s},\boldsymbol{c})$. The inverse of a quantum step is also a quantum step. It is easy to see that:
\begin{equation}
[\hat{T}_\Gamma(\delta,p,\boldsymbol{s},\boldsymbol{c})]^{-1}=\hat{T}_{\Gamma}(\pi+\delta,p,\boldsymbol{c},\boldsymbol{s})
\label{eq:T_inv_definition}
\end{equation}
Since each step of this walk is invertible, the walk $\hat{W}$ itself is invertible. So, given any walk-state there always exist quantum walks that takes it to any of the home-states. Further, we show that it is always possible to do it using exactly $N$ number of $p=1$ quantum steps, where $N=\textbf{max}(\vert b \vert,\vert e \vert)$. 
\subsection{The shrinking algorithm}\label{sub:shrinking_algorithm}
Here we propose an algorithm to find the a walk of $p=1$ quantum steps that can convert the given composite state $\stat{\boldsymbol{U}}$ to a home-state $\stat{\boldsymbol{u}}$.  We start with the simplest case first: when the composite state is of a single term $\stat{\boldsymbol{u};m}$. A quantum step $\hat{T}_\pi(0,1,\boldsymbol{s},\boldsymbol{u}_\perp)$ on $\stat{\boldsymbol{u};m}$ yields $\stat{\boldsymbol{s};m-1}$. Now, from here, taking $m-1$ steps of  $\hat{T}_\pi(0,1,\boldsymbol{s},\boldsymbol{s}_\perp)$, we land on the home-state $\stat{\boldsymbol{s};0}$:\begin{equation}
\stat{\boldsymbol{s};0}={\hat{T}_\pi(0,1,\boldsymbol{s},\boldsymbol{s}_\perp)}^{m-1}\hat{T}_\pi(0,1,\boldsymbol{s},\boldsymbol{u}_\perp)\stat{\boldsymbol{u};m}
\label{eq:Product_to_Home}
\end{equation}We now consider the case of reducing an arbitrary walk-state $\stat{\boldsymbol{U}}$ to some home-state $\stat{\boldsymbol{u};0}$. We achieve using a recursive method: we find a walk step $\hat{T_1}$, that can convert the given state $\stat{\boldsymbol{U}}$ of walk-spread  $[b,e]$ into a composite beam $\stat{\boldsymbol{U}_1}$ of a shorter spread $[b+1,e-1]$. This is always possible if $\stat{\boldsymbol{U}}$  is a walk-state. One set of  parameters of $\hat{T}_1$ that accomplishes this is given by (see appendix):
\begin{equation}
\begin{aligned}
\stat{\boldsymbol{c}}&=\stat{{\boldsymbol{u_e}}_\perp},\\
\Gamma&=2\tan^{-1}\left(\frac{u_e}{u_{e-1}\vert\ip{{\boldsymbol{u}_e}_\perp}{\boldsymbol{u}_{e-1}}\vert}\right),\\
\delta&=\pi-\Phi(\ip{{\boldsymbol{u}_e}_\perp}{\boldsymbol{u}_{e-1}}).
\end{aligned}
\label{eq:shrinking_plate_parameters}
\end{equation}
Now, note that the ensuing state $\stat{\boldsymbol{U}_1}$ is also a walk-state, having obtained through a quantum step on a walk-state. It can also be reduced to another composite state $\stat{\boldsymbol{U}_2}$ whose spread is  $[b+2,e-2]$, by using another walk step $\hat{T}_2$, whose parameters can be computed in a similar manner as in Eq. (\ref{eq:shrinking_plate_parameters}).   Proceeding this way for a total $\frac{e-b}{2}$ steps if  $b$ and $e$ are of same parity, or for $\frac{e-b+1}{2}$ steps if they are of different parity, one lands in a separable state $\stat{\boldsymbol{u};p}$, where $p$ is equal to $\frac{b+e}{2} $ in the former case and  $\frac{b+e-1}{2}$ in the latter case. Now, if $p=0$, we are at home, and therefore have accomplished what we have set out to. Otherwise, transform $\stat{\boldsymbol{u};p}$ into $\stat{\boldsymbol{s};0}$ in $p$ steps as in Eq. (\ref{eq:Product_to_Home}).
Therefore, in either case, in a total of $N=\textbf{max}(\vert b \vert,\vert e \vert)$  steps it is possible to reduce a composite state of spread $[b,e]$ into a home-state:
\begin{equation}
\hat{T}_N\cdots \hat{T}_2\hat{T}_1\stat{\boldsymbol{U}}=\stat{\boldsymbol{u};0}
\end{equation}
The composite state $\stat{\boldsymbol{U}}$ can therefore be constructed from a home-state $\stat{\boldsymbol{u}}$ by taking the inverse of those $N$ steps,  in the opposite order:
\begin{equation}
\stat{\boldsymbol{U}}=\hat{\boldsymbol{W}}_{min} \stat{\boldsymbol{u};0}
\label{eq:mission_statement}
\end{equation}
where 
\begin{equation}
\hat{\boldsymbol{W}}_{min}={\hat{T}_1}^{-1}\cdots{\hat{T}_N}^{-1},
\end{equation}
with the inverse of a walk step $\hat{T}$ given by Eq. (\ref{eq:T_inv_definition}). This completes the algorithm to deterministically reach a given walk-state from some home-state. 

\section{Simplifying the quantum walks \label{sec:Simplifying_QWs}}

\subsection{Simplifying a quantum walk} \label{subsec:simplification}
In Section \ref{sec:Targeted_QWs}, given a walk-state, an algorithm was prescribed for obtaining a walk and a home-state which together generate the given walk-state. Here we discuss a case where the walk $\hat{W}$ of the form Eq. (\ref{eq:Walk_definition}) is given instead of a walk-state. Recall that our definition of the quantum walk allows it to be composed of steps of unequal step-sizes, different coin, shift states etc. Given such a walk, we intend to find another walk  $\hat{W}_{min}$, that functions like $\hat{W}$, but consisting steps of identical step-size $p=1$, and all with the same shift state $\stat{\boldsymbol{s}}$. \\
It is always possible to absorb improper steps $\hat{T}_0(\cdot,\cdot,\boldsymbol{s},\boldsymbol{c})$ into a proper quantum step $\hat{T}_\Gamma\left(\delta,p,\boldsymbol{s},\boldsymbol{c}\right)$ that follows it, as:
\begin{equation}
\hat{T}_\Gamma\left(\delta,p,\boldsymbol{s},\boldsymbol{c}\right)\hat{T}_0\left(\cdot,\cdot,\boldsymbol{p},\boldsymbol{q}\right)=\hat{T}_\Gamma\left(\delta,p,\boldsymbol{s},\boldsymbol{w}\right) 
\label{eq:absorbing_improper_step}
\end{equation}
where the coin-state $\stat{\boldsymbol{w}}$ of the step on the RHS is given by $\stat{\boldsymbol{w}}={\hat{T}^{-1}}_0\left(\cdot,\cdot,\boldsymbol{p},\boldsymbol{q}\right)\stat{\boldsymbol{c}}$. We also have these relations:
\begin{equation}
\begin{aligned}
\hat{T}_\Gamma\left(\delta+2\Delta,p,e^{i\Delta}\boldsymbol{s},e^{i\Delta}\boldsymbol{c}\right)&=	\hat{T}_\Gamma\left(\delta,p,\boldsymbol{s},\boldsymbol{c}\right),\\
\hat{T}_\Gamma\left(-\delta,-p,\boldsymbol{s}_\perp,\boldsymbol{c}_\perp\right)&=	\hat{T}_\Gamma\left(\delta,p,\boldsymbol{s},\boldsymbol{c}\right)
\end{aligned}
\end{equation}
We have the following relation between two quantum steps:
\begin{equation}
\hat{T}_{\Gamma_2}\left(\delta,p,\boldsymbol{s},\boldsymbol{w}\right)\hat{T}_{\Gamma_1}\left(\delta,p,\boldsymbol{w},\boldsymbol{c}\right)\equiv\hat{T}_{\Gamma_1+\Gamma_2}\left(\delta,p,\boldsymbol{s},\boldsymbol{c}\right)
\end{equation}

Given a $\hat{W}$, pick up an arbitrary home-state $\stat{\boldsymbol{p};0}$, and let $\stat{\boldsymbol{P}}$ be the outcome of quantum walk $\hat{W}$ starting with $\stat{\boldsymbol{p};0}$. Now, ignoring $\hat{W}$, we could start from $\stat{\boldsymbol{P}}$ and, from the shrinking algorithm of section \ref{sec:Targeted_QWs}, get a walk $\hat{W}_{s}$, and a  home-state $\stat{\boldsymbol{p}_s}$. Now, how are $\hat{W}_{s}$ and $\hat{W}$ related? Both lead to the same walk-state, but for different home-states $\hat{W}\stat{\boldsymbol{p}}= \hat{W}_{s}\stat{\boldsymbol{p}_s}$.  Both the walks have identical characteristic vector: $\overrightarrow{p}(\hat{W})=\overrightarrow{p}(\hat{W}_{s})$.  Indeed, they are only related by an SU(2) operation in the coin-space:
\begin{equation}
\hat{W} = \hat{W}_{s}\hat{T}_0(\cdot,\cdot,\boldsymbol{p}_s,\boldsymbol{p})
\label{eq:W_Simplified}
\end{equation}
We say the walk on the RHS, $\hat{W}_{s}\hat{T}_0(\cdot,\cdot,\boldsymbol{p}_s,\boldsymbol{p})$, as the walk that ``minimizes the walk $\hat{W}$'', and represent it as  $\textbf{min}\left(\hat{W}\right)$.  The improper step $\hat{T}_0(\cdot,\cdot,\boldsymbol{p}_s,\boldsymbol{p})$ can be absorbed into the first step of $\hat{W}_s$, as in Eq. (\ref{eq:absorbing_improper_step}), so that the minimal walk $\textbf{min}\left(\hat{W}\right)$ contains only as many steps as are in $\hat{W}_s$.
\subsection{Quantum walks between two walk-states} \label{subsec:walk_between_states}
While the paper till now discussed the cases of a quantum walk starting from home-states to the composite walk-states, in this subsection we discuss about going from one composite walk-state to another composite walk-state through a minimal quantum walk. 
That is, given two walk-states $\stat{\boldsymbol{P}}$ and $\stat{\boldsymbol{Q}}$, we seek the minimal quantum walk $\hat{W}_{min}$ such that $\stat{\boldsymbol{Q}}=\hat{W}_{min}\stat{\boldsymbol{P}}$. Note that, by virtue of being $SU(2)$, the walk $\hat{W}_{min}$ must take $\stat{\boldsymbol{P}_\perp}$ to $\stat{\boldsymbol{Q}_\perp}$. One way to generate $\hat{W}_{min}$ is to first obtain some walk $\hat{W}$ that accomplishes this, and then get $\hat{W}_{min}$ as $\textbf{min}\left(\hat{W}\right)$, using the minimization method disccused above. One such walk that takes $\stat{\boldsymbol{P}}$ to $\stat{\boldsymbol{Q}}$ can be generated by making use of the shrinking algorithm as:
\begin{equation}
\hat{W}=\hat{W}_{Q}\hat{T}_0(\cdot,\cdot,\boldsymbol{q},\boldsymbol{p})\hat{W}^{-1}_{P}
\end{equation}
where the walks $\hat{W}_{P}$ and $\hat{W}_{Q}$ are the minimal walks that take the walk-states $\stat{\boldsymbol{P}}$ and $\stat{\boldsymbol{Q}}$ to the home-states $\stat{\boldsymbol{p};0}$ and $\stat{\boldsymbol{q};0}$ respectively, and  $\hat{T}_0(\cdot,\cdot,\boldsymbol{q},\boldsymbol{p})$  is the improper step that takes $\stat{\boldsymbol{p}}$ to $\stat{\boldsymbol{q}}$  in the coin-space: 
\begin{equation*}
\begin{aligned}
\hat{W}_P\stat{\boldsymbol{p};0}&=\stat{\boldsymbol{P}},\\
\hat{W}_Q\stat{\boldsymbol{q};0}&=\stat{\boldsymbol{Q}},\\
\hat{T}_0(\cdot,\cdot,\boldsymbol{q},\boldsymbol{p})&=\left(\stat{\boldsymbol{q}}\bra{\boldsymbol{p}}+\stat{\boldsymbol{q}_\perp}\bra{\boldsymbol{p}_\perp}\right) \otimes I_w
\end{aligned}
\end{equation*}
From these definitions, it is straightforward to verify that $\hat{W}\stat{\boldsymbol{P}}=\stat{\boldsymbol{Q}}$ and $\hat{W}\stat{\boldsymbol{P}_\perp}=\stat{\boldsymbol{Q}_\perp}$, as desired. Note that the quantities appearing on the left-hand side of the above equation, $\hat{W}_P$, $\hat{W}_Q$, $\stat{\boldsymbol{p}}$ and $\stat{\boldsymbol{q}}$, are not available before hand, and must be computed through the shrinking algorithm of Sec (\ref{sec:Targeted_QWs}). 
The minimal quantum walk $\hat{W}_{min}$ to accomplish this  can then be found by minimizing the above walk $\hat{W}$ by the method discussed in subsection \ref{subsec:simplification}.

\section{Illustrations}
\label{sec:Illustrations}
Here we shall provide brief numerical illustrations for the results presented in the earlier sections. 
\subsection{Representing the composite states \label{sec:Representaton}}
\begin{figure*}[htbp]
	\centering
	\includegraphics[height=\textheight,width=\linewidth,keepaspectratio]{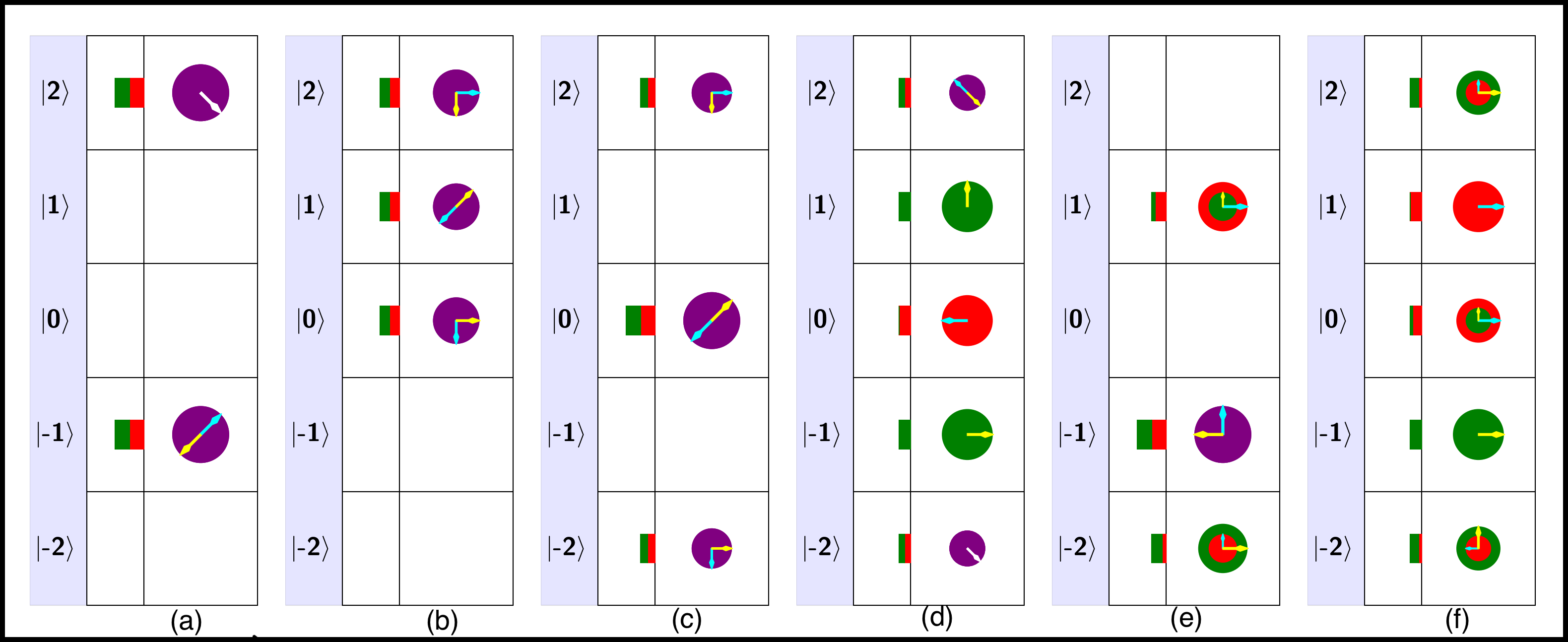}
	\caption{The six composite states of Eq. (\ref{eq:six_states}) depicted in our new notation. The purple circles correspond to the states where the $\stat{\boldsymbol{0}}$ and $\stat{\boldsymbol{1}}$ components are of the same magnitude. The green circle corresponds to the $\stat{\boldsymbol{0}}$ component, its radius proportional to the magnitude of that component and its phase is indicated by an yellow arrow.  The red circle corresponds to the $\stat{\boldsymbol{1}}$ component, its radius being proportional to the magnitude of that component and its phase is indicated by an cyan arrow. }
	\label{fig:All_Six_States}
\end{figure*}

Normalized qubit states of the form Eq. (\ref{eq:qubit_definition}) can be elegantly described on the surface of a unit sphere called the Bloch Sphere (which in the polarization setting is termed as the Poincare sphere \cite{born2013principles}). This treatment, in addition to providing a visualization of quantum states, also aids in giving a geometric picture of the action of SU(2) operators on such quantum states. In the polarization setting, the notion of Poincare sphere has been extended to represent the composite states like Eq. (\ref{eq:arb_composite_state}), but limited to only two terms \cite{padgett1999poincare},\cite{milione2011higher}. A well-known limitation of all these representations has been that they represent a quantum states only up to an overall phase factor. \\ 
In this paper, we introduce a new graphical way of indicating coin-states  $\stat{\boldsymbol{w}}$ of the form Eq. (\ref{eq:qubit_definition}). States of the form $\frac{1}{\sqrt{2}}\left(e^{i\theta_0}\stat{\boldsymbol{0}}+e^{i\theta_1}\stat{\boldsymbol{1}}\right)$ are depicted as a single purple-colored circle, of radius proportional to $\frac{1}{\sqrt{2}}$, with yellow and cyan colored arrows making angles $\theta_0$ and $\theta_1$ with the x-axis respectively. States of the form $\frac{e^{i\theta}}{\sqrt{2}}\left(\stat{\boldsymbol{0}}+\stat{\boldsymbol{1}}\right)$ are depicted with a purple-colored circle of radius proportional to $\frac{1}{\sqrt{2}}$  with a single white-colored arrow making an angle $\theta$ with the x-axis. The rest of the states, we indicate in terms of the two complex numbers, $w_0$ and $w_1$, the component of $\stat{\boldsymbol{w}}$ along  $\stat{\boldsymbol{0}}$ and $\stat{\boldsymbol{1}}$ unit vectors respectively. We represent these complex numbers as two concentric green and red circles of radii proportional to $\abs{w_0}$ and $\abs{w_1}$ respectively. The phase of these complex numbers are indicated by yellow and cyan colored arrows over these circles, the angles they make with the horizontal indicative of their phases.  \\
This representation can be carried over to the composite states as well. We depict them as follows: on the y-axis we mark the walk-space basis states $\stat{m}$ from $m\in [b,e]$. At every $m$ with non-zero $u_m$, the corresponding coin-state $\stat{\boldsymbol{u}_m}$ is depicted as described above, except that the radii of these circles is scaled by the amplitude $u_m$. 
This representation of composite states, while by no means as versatile as that of Poincare Sphere, will nevertheless serve us well to better illustrate the results derived earlier in the manuscript. 
\subsection{Walk and non-walk states}
\label{subsec:walk_non_walk}
Consider as examples the following six composite states: \begin{equation}
\begin{aligned}
\stat{\boldsymbol{A}}&=\frac{1}{\sqrt{2}}\left(\stat{\boldsymbol{a},-1}+\stat{\boldsymbol{d},2}\right),\\
\stat{\boldsymbol{B}}&=\frac{1}{\sqrt{3}}\left(\stat{\boldsymbol{h},0}-\stat{\boldsymbol{a},1}+\stat{\boldsymbol{v},2}\right),\\
\stat{\boldsymbol{C}}&=\frac{1}{2}\left(\stat{\boldsymbol{h},-2}-\sqrt{2}\stat{\boldsymbol{a},0}+\stat{\boldsymbol{v},2}\right),\\
\stat{\boldsymbol{D}}&=\frac{1}{\sqrt{5}}\left(\stat{\boldsymbol{d};-2}+\stat{\boldsymbol{l};-1}-\stat{\boldsymbol{r};0}+i\stat{\boldsymbol{l};1}+i\stat{\boldsymbol{a};2}\right),\\
\stat{\boldsymbol{E}}&=\frac{1}{2}\left(\stat{\boldsymbol{c},-2}-\sqrt{2}\stat{\boldsymbol{h},-1}+\stat{\boldsymbol{e},1}\right),\\
\stat{\boldsymbol{F}}&=\frac{1}{\sqrt{5}}\left(i\stat{\boldsymbol{c};-2}+\stat{\boldsymbol{l};-1}+\stat{\boldsymbol{e};0}+\stat{\boldsymbol{r};1}+\stat{\boldsymbol{c};2}\right)
\end{aligned}
\label{eq:six_states}
\end{equation}
where $\stat{\boldsymbol{h}},\stat{\boldsymbol{v}}$ are a pair of orthogonal states  $\stat{\boldsymbol{h}}=\frac{1}{\sqrt{2}}\left(\stat{\boldsymbol{0}}-i\stat{\boldsymbol{1}}\right)$, and  $\stat{\boldsymbol{v}}=\frac{1}{\sqrt{2}}\left(\stat{\boldsymbol{1}}-i\stat{\boldsymbol{0}}\right)$, and states $\stat{\boldsymbol{d}},\stat{\boldsymbol{a}}$ are another pair of orthogonal states: $\stat{\boldsymbol{d}}=\frac{e^{-i\frac{\pi}{4}}}{\sqrt{2}}\left(\stat{\boldsymbol{1}}+\stat{\boldsymbol{0}}\right)$, and $\stat{\boldsymbol{a}}=\frac{e^{i\frac{\pi}{4}}}{\sqrt{2}}\left(\stat{\boldsymbol{1}}-\stat{\boldsymbol{0}}\right)$, and $\stat{\boldsymbol{c}}$ and $\stat{\boldsymbol{e}}$ are yet another orthogonal pair $\stat{\boldsymbol{c}}=\frac{1}{2}\left(\sqrt{3}\stat{\boldsymbol{0}}+i\stat{\boldsymbol{1}}\right)$ and  $\stat{\boldsymbol{e}}=\frac{1}{2}\left(i\stat{\boldsymbol{0}}+\sqrt{3}\stat{\boldsymbol{1}}\right)$, all belonging to the coin-space. States $\stat{\boldsymbol{D}}$ and $\stat{\boldsymbol{F}}$ are five-component composite states having equal contribution $P_m=\frac{1}{5}$ in each of the positions: $m\in[-2,-1,0,1,2]$.  All these states are normalized. Of these, the first state has the position span of [-1,2], the second state, $\stat{\boldsymbol{B}}$ has the span [0,2], and the rest of the states have the span [-2,2]. 
Of the six states, it can be seen that only the first four states are walk-states satisfying the constraints Eq. (\ref{eq:TI_Constraints}). 
The six states are depicted in Fig. (\ref{fig:All_Six_States}). 
\subsection{Representing quantum steps and quantum walks}
\begin{figure*}[htbp]
	\centering
	\includegraphics[height=\textheight,width=\linewidth,keepaspectratio]{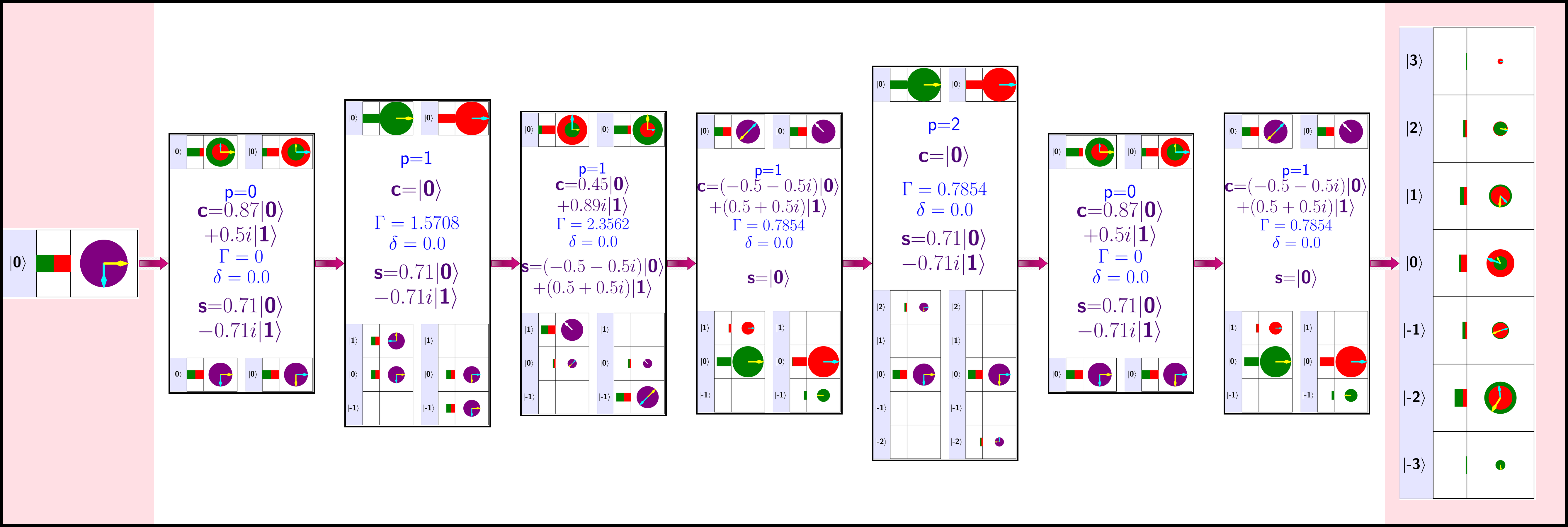}
	\caption{A quantum walk of seven steps, starting from the home-state $\stat{\boldsymbol{h};0}$. At the end of the seventh step, the resulting composite state is the one spanning the positions [-3,3], with probabilities $P_m$ being [0.021,0.355,0.117,0.214,0.197,0.089,0.007] for $m=-3,\cdots,3.$}
	\label{fig:evolution_demo}
\end{figure*}

We represent quantum steps $\hat{T}_\Gamma\left(\delta,p,\boldsymbol{s},\boldsymbol{c}\right)$, in terms of their action on the orthogonal pair of home-states $\stat{\boldsymbol{c};0}$ and $\stat{\boldsymbol{c}_\perp;0}$. The two input home-states and the two output composite states are all depicted in the concentric circle representation of states introduced earlier. A quantum walk $\hat{W}$ of the form Eq. (\ref{eq:Walk_definition}) is represented with its steps stacked from the left to right, with the first step appearing left-most and the last step appearing right-most. 
As an illustration, a quantum walk of seven steps is depicted in Fig. (\ref{fig:evolution_demo}). The first and six steps of this walk are improper steps, the fifth step is of step-size $p=2$, while the rest of the steps are of unit step-size.  
\subsection{Constructing new walks from existing ones}
We now provide a couple of illustrations for constructing new walk-states of larger span from linear combination of walk-states with smaller span. Consider the five-component state $\stat{\boldsymbol{D}}$ of Eq. (\ref{eq:six_states}). Its position span is $[-2,2]$. The position span of $\stat{\boldsymbol{D}_{+3}}$ is therefore $[1,5]$. Its orthogonal state $\stat{(\boldsymbol{D}_{+3})_\perp}$ therefore has a span of $[-5,-1]$. A complex combination of these two states $\alpha\stat{\boldsymbol{D}_{+3}}+\beta\stat{(\boldsymbol{D}_{+3})_\perp}$ where $\abs{\alpha}^2+\abs{\beta}^2=1$ will be walk-states with position span $[-5,5]$. From this state, a ten-term walk-state $\stat{\boldsymbol{G}}$  can be generated as:  
\begin{equation}
\begin{aligned}
\stat{\boldsymbol{G}}=&\frac{1}{\sqrt{2}}\left( \stat{\boldsymbol{D}_{+3}}+\stat{(\boldsymbol{D}_{+3})_\perp}\right)\\
=&\frac{1}{\sqrt{10}}(i\stat{\boldsymbol{d};-5}-i\stat{\boldsymbol{r};-4}+\stat{\boldsymbol{l};-3}+\stat{\boldsymbol{r};-2}+\stat{\boldsymbol{a};-1}\\+&\stat{\boldsymbol{d};1}+\stat{\boldsymbol{l};2}-\stat{\boldsymbol{r};3}+i\stat{\boldsymbol{l};4}+i\stat{\boldsymbol{a};5})
\end{aligned}
\label{eq:state_G}
\end{equation}State $\stat{\boldsymbol{G}}$ has the position span of $[-5,5]$, in which for all non-zero $m \in[-5,5]$, the probability $P_m$ is $\frac{1}{10}$. 
One could also generate a ten-term  ``uniform walk-state'' from $\stat{\boldsymbol{D}}$ as: \begin{equation}
\begin{aligned}
\stat{\boldsymbol{H}}=& \frac{1}{\sqrt{2}}\left(\stat{\boldsymbol{D}_{\times 2}}+\stat{((\boldsymbol{D}_{\times 2})_\perp)_{+1}}\right)\\
=&\frac{1}{\sqrt{10}}(\stat{\boldsymbol{d};-4}+i\stat{\boldsymbol{d};-3}+\stat{\boldsymbol{l};-2}-i\stat{\boldsymbol{r};-1}-\stat{\boldsymbol{r};0}\\+&\stat{\boldsymbol{l};1}+i\stat{\boldsymbol{l};2}+\stat{\boldsymbol{r};3}+i\stat{\boldsymbol{a};4}+\stat{\boldsymbol{a};5}),
\end{aligned}
\label{eq:state_H}
\end{equation} where $\stat{\boldsymbol{D}_{\times 2}}$ is the state that is obtained from $\stat{\boldsymbol{D}}$ by Eq. (\ref{eq:U_times_d}), state $\stat{(\boldsymbol{D}_{\times 2})_\perp}$  is its orthogonal state, and the state $\stat{((\boldsymbol{D}_{\times 2})_\perp)_{+1}}$ is $\stat{(\boldsymbol{D}_{\times 2})_\perp}$ translated by one unit. State $\stat{\boldsymbol{H}}$ has the position span of $[-4,5]$, in which for all $m \in[-4,5]$, including $m=0$, the occupation probability $P_m$ is $\frac{1}{10}$. \\It is easy to see that starting from a home-state, a similar procedure can be applied to construct uniform walk-states of any span $[b,e]$, with $P_m=\frac{1}{e-b+1}$ for all $m\in[b,e]$.
\subsection{Simplification of quantum walks} 
\begin{figure}[htbp]
	\centering
	\includegraphics[height=\textheight,width=\linewidth,keepaspectratio]{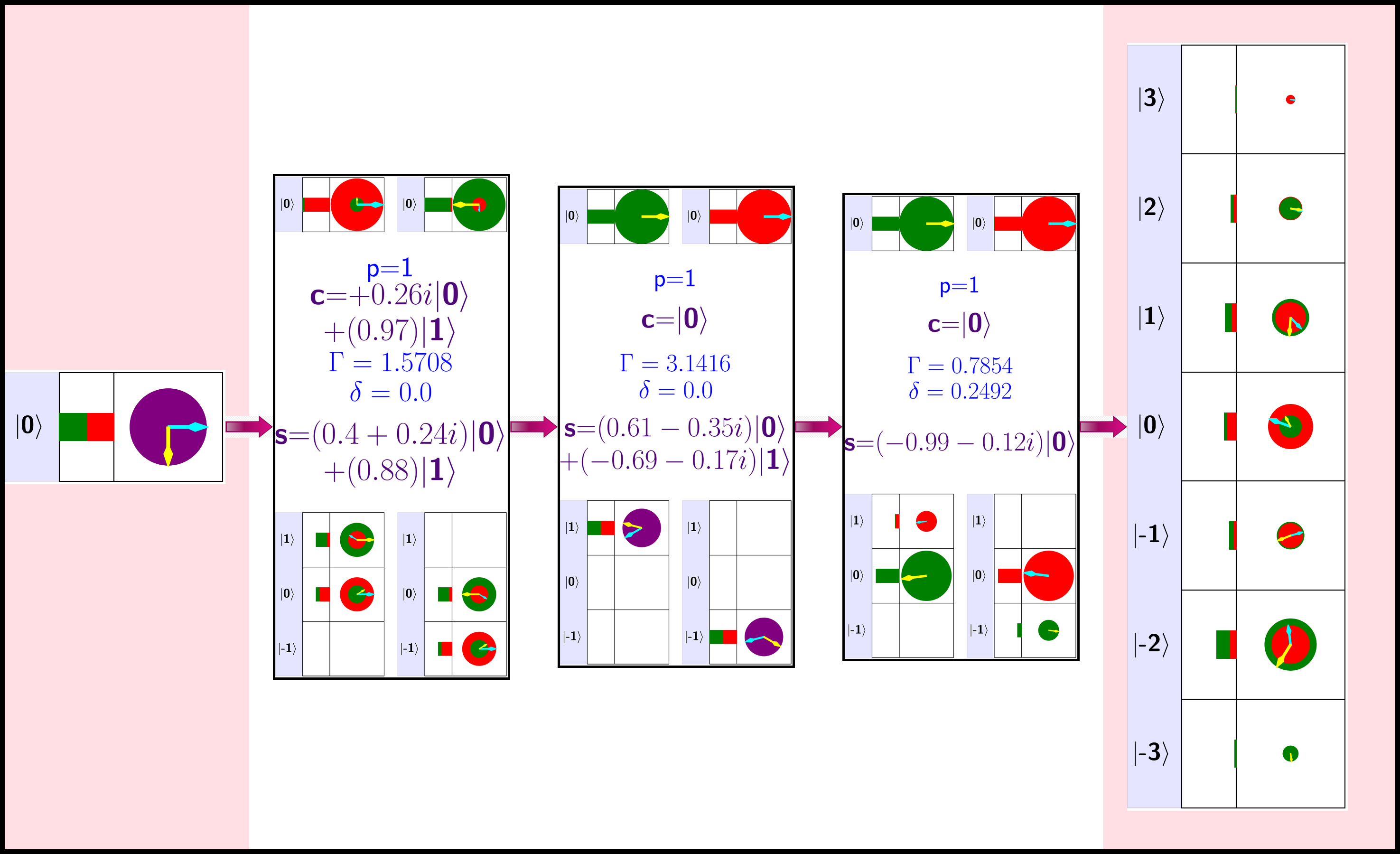}
	\caption{The three-step quantum walk that is equivalent to the seven-step quantum walk depicted in Fig. (\ref{fig:evolution_demo}). The initial and final are the same as those in Fig. (\ref{fig:evolution_demo}).}
	\label{fig:equivalent_walk}
\end{figure}

For an illustration of the simplification of a quantum walk $\hat{W}$ discussed in subsection \ref{subsec:simplification}, consider the seven-step quantum walk depicted in Fig. (\ref{fig:evolution_demo}). Recall that of the seven steps there, two were improper, that is, pure SU(2) transformations in the coin-space, one step was of $p=2$ , and remaining four steps were of $p=1$. The position span of the emerging composite state is $[-3,3]$. This state can therefore be obtained from the home-state in just three $p=1$ steps, using the shrinking algorithm. This three-step minimal walk is depicted in Fig. (\ref{fig:equivalent_walk}).

\subsection{Illustrations of the shrinking algorithm}

\begin{figure*}[tbp]
	\centering
	\includegraphics[height=\textheight,width=\linewidth,keepaspectratio]{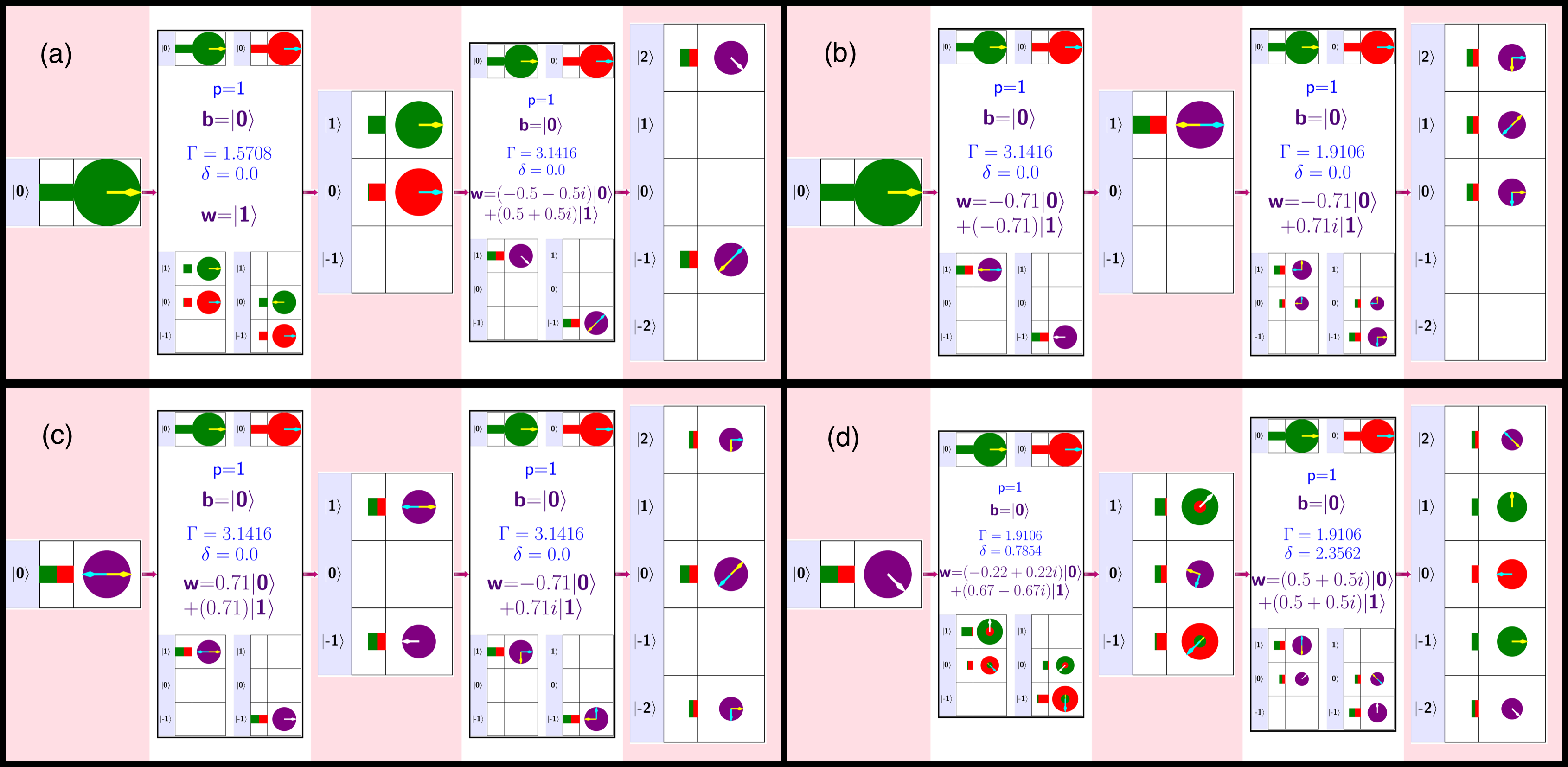}
	\caption{Targeted quantum walks to generate each of the four walk-states of Eq. (\ref{eq:six_states}) starting from home-states, in two steps of unit step size. Figures (a),(b),(c) and (d) corresponding to the states $\stat{\boldsymbol{A}},\stat{\boldsymbol{B}},\stat{\boldsymbol{C}}$ and $\stat{\boldsymbol{D}}$ respectively.}
	\label{fig:four_states_targeted_walk}
\end{figure*}
As illustration of the shrinking algorithm of subsection (\ref{sub:shrinking_algorithm}), we show the evolution of home-states into each of the four walk-states of Eq. (\ref{eq:six_states}). The characteristic vector (defined in Eq. (\ref{eq:characteristic_function})) of these four walk-states are $\overrightarrow{p}(\boldsymbol{A})=(0,\frac{1}{2},\frac{1}{2}), \overrightarrow{p}(\boldsymbol{B})=(\frac{1}{3},\frac{1}{3},\frac{1}{3}), \overrightarrow{p}(\boldsymbol{C})=(\frac{1}{2},0,\frac{1}{2})$ and $\overrightarrow{p}(\boldsymbol{D})=(\frac{1}{5},\frac{2}{5},\frac{1}{5})$. From these, it can be inferred that no pair of them can be obtained from a single quantum walk acting on different home-states.  As $N=2$ for all four states, they can be generated in just two proper steps.
The required steps, the initial home-states and the intermediate walk-states are all depicted in Fig. (\ref{fig:four_states_targeted_walk}). While the states $\stat{\boldsymbol{A}}$ and $\stat{\boldsymbol{B}}$ are generated with $\stat{\boldsymbol{l};0}$ as the home-state, the other two are generated with home-states  $e^{i\frac{3\pi}{4}}\stat{\boldsymbol{a};0}$ and $\stat{\boldsymbol{d};0}$ respectively. 
\subsection{Transformation between walk-states}
\begin{figure*}[htbp]
	\centering
	\includegraphics[height=\textheight,width=\linewidth,keepaspectratio]{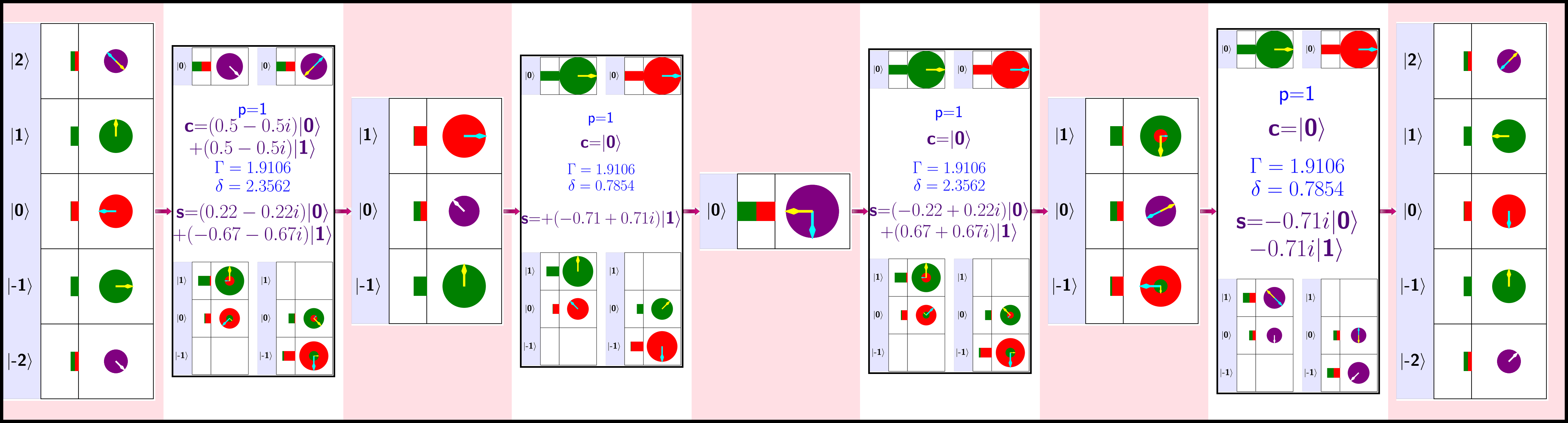}
	\caption{A minimal quantum walk for transforming the state  $\stat{\boldsymbol{D}}$ of Eq. (\ref{eq:six_states}) to the state $i\stat{\boldsymbol{D}}$. The walk is four step long.}
	\label{fig:state_D_Phase}
\end{figure*}

As an illustration of transforming one walk-state to another as detailed in subsection \ref{subsec:walk_between_states}, consider the transformation from some state walk-state $\stat{\boldsymbol{U}}$ to a walk-state $e^{i\Delta}\stat{\boldsymbol{U}}$, where $\Delta$ is some phase factor, $0\le\Delta < 2\pi$. The two states differ only by an overall phase factor. Despite of that, the quantum walk $\hat{W}$ that accomplishes the transformation, $\hat{W}\stat{\boldsymbol{U}}=e^{i\Delta}\stat{\boldsymbol{U}}$, is far from trivial, unless $\Delta=0\text{ or }\pi$. This can be understood as follows: $\stat{\boldsymbol{U}}$ can be taken to $e^{i\Delta}\stat{\boldsymbol{U}}$ by transforming each constituent term $\stat{\boldsymbol{u}_m}$ of $\stat{\boldsymbol{U}}$ to $e^{i\Delta}\stat{\boldsymbol{u}_m}$. This cannot be accomplished by SU(2) transformation in the coin-space alone, as the transformation that takes $\stat{\boldsymbol{u}_m}$ to $e^{i\Delta}\stat{\boldsymbol{u}_m}$ will not take $\stat{\boldsymbol{u}_n}$ to $e^{i\Delta}\stat{\boldsymbol{u}_n}$, unless $\stat{\boldsymbol{u}_m}$ and $\stat{\boldsymbol{u}_n}$ themselves differ by an overall phase. One, therefore, has to take recourse to a multi-step quantum walk only. \\
Consider, for instance, the transformation of the state $\stat{\boldsymbol{D}}$ of Eq. (\ref{eq:six_states}), to the walk-state $i\stat{\boldsymbol{D}}$. The walk that accomplishes this can be obtained by the minimization procedure elaborated above. The four step minimal quantum walk is shown in Fig. (\ref{fig:state_D_Phase}). The characteristic vector of this walk is $(0.04,0.16,0.48,0.16,0.16)$. The minimal quantum walk for realizing $\hat{W}^2$ on the other hand, is, as expected, just an SU(2) operator in the coin-space, $-\hat{I}_2$, the negative of identity operator. 

\section{Optical implementation }
\label{sec:Optical_Implementation}
We shall now provide a means of realizing the aforementioned ideas on an actual physical hardware over which such quantum walks are being implemented. In the linear optical implementation of DTQWs, the spin angular momentum (SAM) of the light beam as the coin-space $H_c$, and its orbital angular momentum (OAM) acts as the walk space, $H_w$. The state of polarization (SoP) of a light beam characterizes the SAM of light, with left circular polarization identified with $+1$ units of SAM and right-circular polarization identified with $-1$ units of it. OAM, on the other hand, is independent of the SAM of the light beam and characterizes the helicity of its wavefront. An arbitrary composite state of the light beam is the one which exists in a superposition of combinations of spin and orbital angular momenta, and a definite SoP or OAM cannot be assigned to them. Such beams are termed vector beams, and they can be identified by the varying phase and SoP distribution across their transverse plane. 
\subsection{Scalar and vector beams}
An arbitrary SoP of a light beam can be represented as a two-component unit vector, $\stat{\boldsymbol{u}}$ called the ``Jones vector'', in some orthogonal basis, as in Eq. (\ref{eq:qubit_definition}). 
We identify the states $\stat{\boldsymbol{0}}$ and $\stat{\boldsymbol{1}}$ with $\stat{\boldsymbol{l}}$ and $\stat{\boldsymbol{r}}$, the left and right circular polarization states of light respectively. The states $\stat{\boldsymbol{h}}$,$\stat{\boldsymbol{v}}$, $\stat{\boldsymbol{d}}$ and $\stat{\boldsymbol{a}}$, defined following Eq. (\ref{eq:six_states}), will then correspond to horizontal, vertical, diagonal and anti-diagonal SoPs of light, respectively.
In our numerical simulations, the  $\stat{\boldsymbol{h}}$ and $\stat{\boldsymbol{v}}$  states correspond to the standard basis vectors $[1,0]^T$ and $[0,1]^T$ respectively. The left and right circular SoPs $\stat{\boldsymbol{l}}$ and $\stat{\boldsymbol{r}}$ would then correspond to the column vectors $[1,i]^T$ and $[i,1]^T$ respectively. \\
Unitary transformations between pairs of SoPs is affected using waveplates. In this paper, we represent them by the symbol $\hat{p}_\Gamma(\alpha)$, where $\Gamma$ is its retardance, and $\alpha$ is the angle its fast-axis makes with the x-axis. The action of this waveplate on an arbitrary state of polarization $\stat{\boldsymbol{u}}$ is given by:
\begin{equation}
	\begin{aligned}
		\hat{p}_\Gamma(\alpha)\stat{\boldsymbol{u}}=&-\sin\frac{\Gamma}{2}\ip{\boldsymbol{r}}{\boldsymbol{u}}e^{-2i\alpha}\stat{\boldsymbol{l}}\\&+\cos\frac{\Gamma}{2}\stat{\boldsymbol{u}}+\sin\frac{\Gamma}{2}\ip{\boldsymbol{l}}{\boldsymbol{u}}e^{2i\alpha}\stat{\boldsymbol{r}}
	\end{aligned}
	\label{eq:action_waveplate}
\end{equation}
Here $\ip{\boldsymbol{l}}{\boldsymbol{u}}$ and $\ip{\boldsymbol{r}}{\boldsymbol{u}}$ stand for the left and right circular polarization components of $\stat{\boldsymbol{u}}$. 
It is important to note it is always possible to transform one SOP to another SOP ``up to a phase" using a single waveplate. That is: given a pair of SoPs $\stat{\boldsymbol{u}}$ and $\stat{\boldsymbol{s}}$, there exists a waveplate $\hat{p}_\Gamma(\alpha)$ such that $\hat{p}_\Gamma(\alpha)\stat{\boldsymbol{u}}= e^{i\delta}\stat{\boldsymbol{s}} $ where $\delta$ is an extra phase factor. The $\Gamma, \alpha$ of the waveplate needed for this transformation, and the additional phase factor $\delta$ generated in this process, can be derived from Eq. (\ref{eq:action_waveplate}). Furthermore, it always possible to transform $\stat{\boldsymbol{u}}$ to $\stat{\boldsymbol{s}}$, without any additional phase factors, by using a three-plate gadget consisting of a half-wave plate inserted in between two quarter-wave plates (\cite{SIMON1989474,simon1990minimal}). 

\begin{figure}[htbp]
	\centering
	\includegraphics[height=\textheight,width=\linewidth,keepaspectratio]{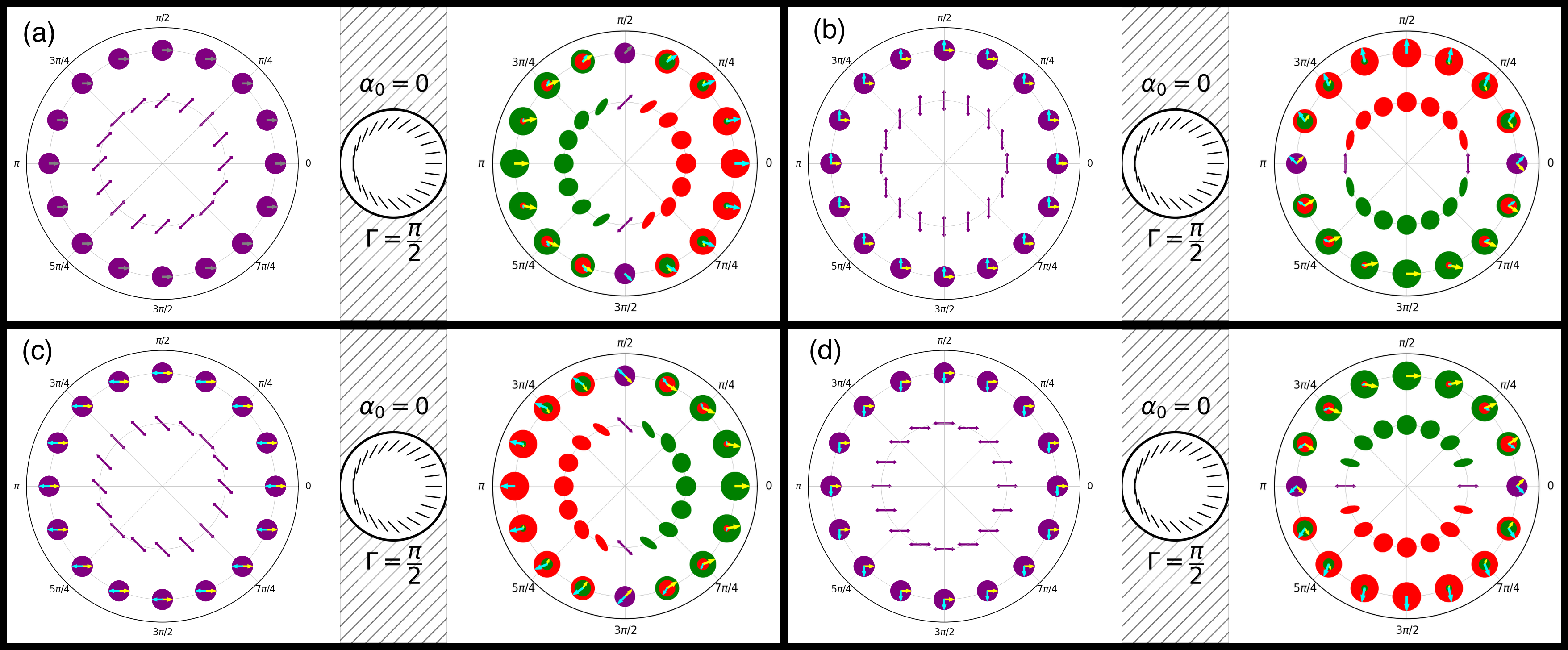}
	\caption{(a-d) Action of a $q=\frac{1}{2}$ q-plate with $\Gamma=\frac{\pi}{2}$ and $\alpha=0$, on four linearly polarized light beams, $\stat{\boldsymbol{d}},\stat{\boldsymbol{v}}\stat{\boldsymbol{a}}$ and $\stat{\boldsymbol{h}}$ respectively, having zero OAM. The input and output beams are represented in terms of the concentric-circles notation introduced earlier and also in the standard notation of the polarization ellipses.}
	\label{fig:OAM_Gamma_pi_2_q_plate}
\end{figure}

\begin{figure}[htbp]
	\centering
	\includegraphics[height=\textheight,width=\linewidth,keepaspectratio]{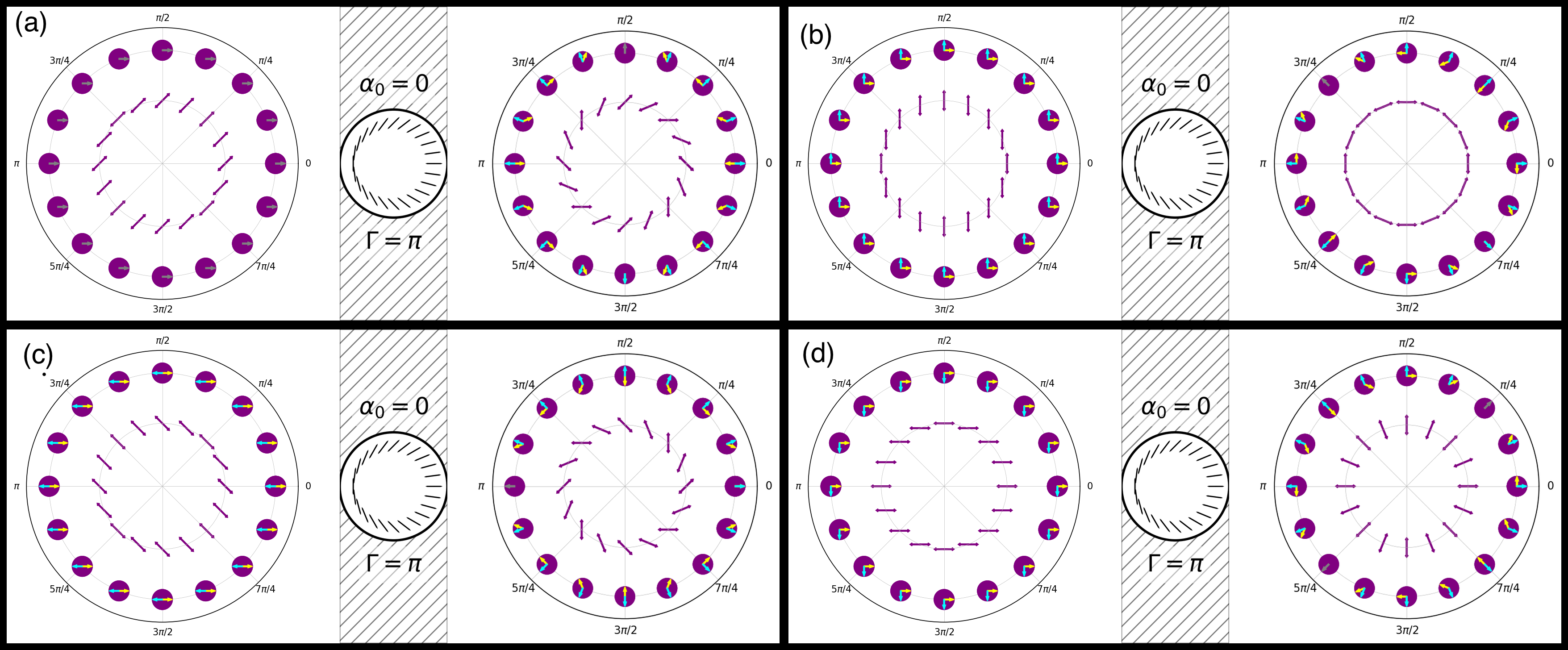}
	\caption{(a-d) Action of a $q=\frac{1}{2}$ q-plate with $\Gamma=\pi$ and $\alpha=0$, on four linearly polarized light beams, $\stat{\boldsymbol{d}},\stat{\boldsymbol{v}}\stat{\boldsymbol{a}}$ and $\stat{\boldsymbol{h}}$ respectively, having zero OAM. }
	\label{fig:OAM_Gamma_pi_q_plate}
\end{figure}

A vector beam is a light beam whose SoP varies across its transverse plane. In this work, we shall confine our attention to only cases where the variation is only along the azimuthal direction. We therefore ignore the radial co-ordinate in the rest of our discussion. The SoP of a vector beam will be represented as $\stat{\boldsymbol{u}(\varphi)}$ where $\varphi$ is the azimuthal angle, measured from some reference axis, say the x-axis. It can be represented as in Eq. (\ref{eq:qubit_definition}):
\begin{equation}
	\stat{\boldsymbol{u}(\varphi)}=u_l(\varphi)\stat{\boldsymbol{l}}+u_r(\varphi)\stat{\boldsymbol{r}}
	\label{eq:Vector_Beam_SoP_definition}
\end{equation} where $u_l(\varphi)$ and $u_r(\varphi)$ are complex functions of the azimuthal angle $\varphi$, satisfying the normalization requirement $\vert u_l(\varphi)\vert ^2+\vert u_r(\varphi)\vert ^2=1$. 
If the coefficients $u_l(\varphi)$ and $u_r(\varphi)$ depend identically on $\varphi$ as $e^{im\varphi}$, for some integer $m$, we say the light beam carries an OAM of $m$ units. We represent such states by the symbol $\stat{\boldsymbol{u};m}$:
\begin{equation}
	\stat{\boldsymbol{u};m} =\stat{\boldsymbol{u}}\otimes\stat{m}=e^{im\varphi}\stat{\boldsymbol{u}}
	\label{eq:separable}
\end{equation}
where $m$ is an integer and $\stat{\boldsymbol{u}}$ is an SoP of the form Eq. (\ref{eq:qubit_definition}), having no azimuthal dependence. We refer to such light beams as scalar light beams. The SoPs along two different azimuthal angles on the transverse plane of such scalar light beams differ only in their overall global phase. An arbitrary vector beam of the form Eq. (\ref{eq:Vector_Beam_SoP_definition}) can be decomposed as a superposition of a collection of such scalar beams, resulting in a composite state $\stat{\boldsymbol{W}}$ of the form Eq. (\ref{eq:arb_composite_state}).  The Jones vector of its $m^{th}$ OAM component and its amplitude can be extracted from $\stat{\boldsymbol{u}(\varphi)}$ as 
\begin{equation}
	u_m\stat{\boldsymbol{u}_m} =\int_{0}^{2\pi}e^{-im\varphi}\stat{\boldsymbol{u}(\varphi)}d\varphi,\forall m\in\left[b,e\right]
	\label{eq:OAM_Component_Extraction}
\end{equation}
The limits $b$ and $e$ in Eq. (\ref{eq:OAM_Component_Extraction}) are the smallest and largest $m$ that yield non-zero $u_m$. We shall confine our attention to only those $\boldsymbol{u}(\varphi)$ for which the $b$ and $e$ are finite, i.e., only those vector beams which are constructed out of a superposition of a finite number of scalar beams. In such a case, there is a one-to-one correspondence between vector beams having a spatially SoP of the form Eq. (\ref{eq:Vector_Beam_SoP_definition}) and composite quantum state of the form Eq. (\ref{eq:arb_composite_state}), $\stat{\boldsymbol{u}(\varphi)}\leftrightarrow\stat{\boldsymbol{U}}$, with Eq. (\ref{eq:OAM_Component_Extraction}) establishing the mapping. The corresponding orthogonal states also follow the same mapping $\stat{{\boldsymbol{u}(\varphi)}_\perp}\leftrightarrow\stat{\boldsymbol{U}_\perp}$, where $\stat{{\boldsymbol{u}(\varphi)}_\perp}$ is obtained through $\stat{{\boldsymbol{u}(\varphi)}}$ from Eq. (\ref{eq:qubit_perp}), and $\stat{\boldsymbol{U}_\perp}$ is obtained through $\stat{\boldsymbol{U}}$ using Eq. (\ref{eq:U_perp}). \\A prominent method of generating such vector light beams has been using optical interferometers\cite{liu2012generation,li2016generation} with spatial light modulators\cite{rosales2017simultaneous,chen2015complete,gao2019single,garcia2020efficient}. They have also been generated by employing inhomogenous waveplates called q-plates \cite{Rubano:19,rubano2019q} which are waveplates with uniform retardance, but whose orientation of the fast axis varies linearly with the azimuthal angle. We represent such q-plates with the symbol $\hat{q}_\Gamma\left(q,\alpha_0\right)$, where $\Gamma$ is called the retardance, $q $ is called the topological charge, and $\alpha_0$ is the offset-angle \cite{kadiri2019wavelength}. The import of these parameters can be understood by knowing its action on an arbitrary separable state $\stat{\boldsymbol{u};m}$:
\begin{equation}
	\begin{aligned}
		\hat{q}_\Gamma\left(q,\alpha_0\right)\stat{\boldsymbol{u};m}=&-\sin\frac{\Gamma}{2}\ip{\boldsymbol{r}}{\boldsymbol{u}}e^{-2i\alpha_0}\stat{\boldsymbol{l};m-2q}\\&+\cos\frac{\Gamma}{2}\stat{\boldsymbol{u};m}+\sin\frac{\Gamma}{2}\ip{\boldsymbol{l}}{\boldsymbol{u}}e^{2i\alpha_0}\stat{\boldsymbol{r};m+2q}
	\end{aligned}
	\label{eq:q_mu_plate_definition}
\end{equation}A q-plate of topological charge $q$ and retardance $\pi$ is called as the standard q-plate. It raises the OAM of the left circular component of the input light beam by $2q$ units and simultaneously reduces the OAM of its right circular component by $2q$ units. The inverses of $\hat{p}_\Gamma(\alpha)$ and $\hat{q}_\Gamma\left(q,\alpha_0\right)$ are given by 
\begin{equation}
\begin{aligned}
	\hat{p}_\Gamma(\alpha)^{-1}&=\hat{p}_\Gamma(\alpha+\frac{\pi}{2}),\\
	\hat{q}_\Gamma\left(q,\alpha_0\right)^{-1}&=\hat{q}_\Gamma\left(q,\alpha_0+\frac{\pi}{2}\right)
\end{aligned}
\end{equation}
A standard method for representing an SoP has been as polarization ellipses \cite{collett2003polarized}. In this notation, the SoPs are depicted as ellipses, with ellipsisity $e$ and orientation $\phi$ given by:
\begin{equation}
\begin{aligned}
e&=\abs{\abs{\ip{\boldsymbol{l}}{\boldsymbol{u}}}^2-\abs{\ip{\boldsymbol{r}}{\boldsymbol{u}}}^2},\\
\phi&=\Phi\left(\ip{\boldsymbol{r}}{\boldsymbol{u}}\right)-\Phi\left(\ip{\boldsymbol{l}}{\boldsymbol{u}}\right)+\frac{\pi}{4}.
\end{aligned}
\end{equation}
These definitions are not standard, but have been employed here due to their simplicity. The color of the ellipse indicates the helicity of SoP: states with $\abs{\ip{\boldsymbol{l}}{\boldsymbol{u}}}>\abs{\ip{\boldsymbol{r}}{\boldsymbol{u}}}$) are colored green and those with $\abs{\ip{\boldsymbol{l}}{\boldsymbol{u}}}<\abs{\ip{\boldsymbol{r}}{\boldsymbol{u}}}$ are colored red. States with $\abs{\ip{\boldsymbol{l}}{\boldsymbol{u}}}=\abs{\ip{\boldsymbol{r}}{\boldsymbol{u}}}$  are depicted as purple-colored arrows. Recall that the polarization ellipses representation depicts an SoP only upto a phase.\\
In this paper, the azimuthally-varying SoP of the vector beams is depicted in both the methods: (i) as polarization ellipses and (ii) as concentric circles introduced in subsection (\ref{sec:Representaton}). 
To map the two representations: the circular polarizations are represented by green and red circles in both the representations. Purple circles in the concentric-circles representation corresponds to purple arrows in the elliptical representation. \\
In Figs. (\ref{fig:OAM_Gamma_pi_2_q_plate}) and (\ref{fig:OAM_Gamma_pi_q_plate}), we provide a quick overview of the action of $\Gamma=\frac{\pi}{2}$ and $\Gamma=\pi$ q-plates on four linearly-polarized scalar light beams. These serve not only in giving a quick recall of the functioning of q-plates, but also in connecting our new notation of representing the SoPs, with the standard polarization ellipses representation.  

\subsection{Discrete time quantum walks on the SAM-OAM space}
\begin{figure*}[htbp]
	\centering
	\includegraphics[height=\textheight,width=\linewidth,keepaspectratio]{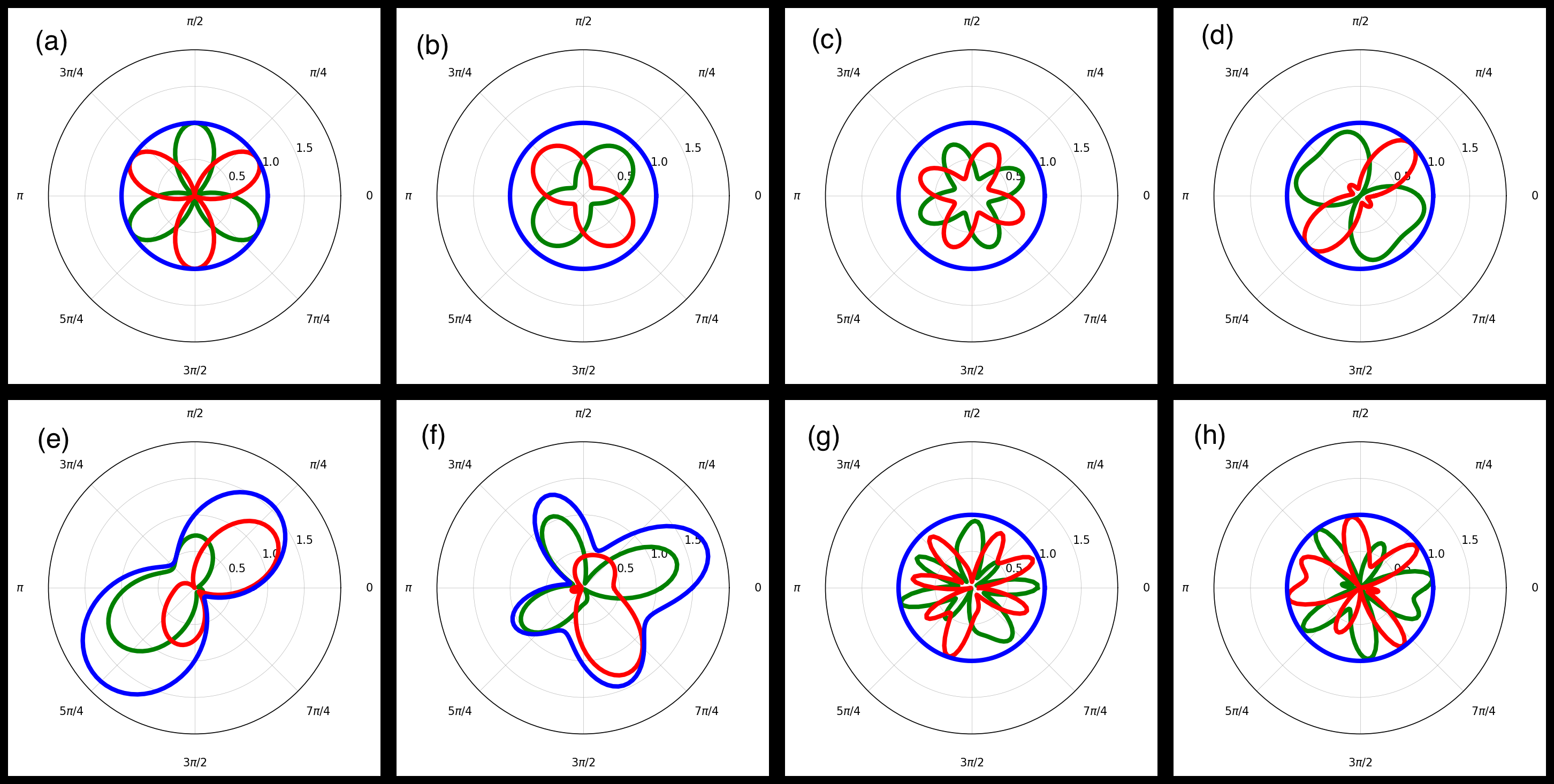}
	\caption{Polar plot the intensity distribution of the six composite states of Eq. (\ref{eq:six_states}), state $\stat{\boldsymbol{G}}$ of Eq. (\ref{eq:state_G}) and $\stat{\boldsymbol{H}}$ of Eq. (\ref{eq:state_H}), as a function of the azimuthal angle. The green and red plots correspond to the intensity in the left and right circular polarization components, while the black plot correspond to the total, that is the sum of these intensities.}
	\label{fig:OAM_eight_intensities}
\end{figure*}

  In the SAM-OAM implementation of quantum walks, the ``coin-toss operation'' is implemented using waveplates and the ``shift operation'' is implemented using q-plates \cite{zhang2010implementation,goyal2013implementing}. 
Since the intial state is normalized, and the involved q-plates and waveplates are all unitary, an immediate constraint on the emerging vector beams $\stat{\boldsymbol{U}}$ is that $\ip{\boldsymbol{U}}{\boldsymbol{U}}=1$ at all azimuthal angles $\varphi$:
\begin{equation}
	\ip{\boldsymbol{u}(\varphi)}{\boldsymbol{u}(\varphi)}=\ip{\boldsymbol{U}}{\boldsymbol{U}}=\sum_{m,n}^{}u_mu_n\ip{\boldsymbol{u}_m}{\boldsymbol{u}_n}e^{i(m-n)\varphi}=1
	\label{eq:oam_constraints}
\end{equation}
This seemingly innocuous normalization requirement leads to $e-b+1$ constraints on the constituent polarizations and OAM amplitudes, which are exactly identical to that of Eq. (\ref{eq:TI_Constraints}). Therefore, this leads us to conclude that all normalized composite states in the SAM-OAM space of the light beams are walk-states.  
For instance, the vector beams corresponding to the six states of Eq. (\ref{eq:six_states}) are depicted in Fig. (\ref{fig:OAM_eight_intensities}). In each of these eight plots, we have plotted the right circular component $\vert \ip{\boldsymbol{r}}{\boldsymbol{U}}\vert ^2$ in red line, the left circular component $\vert \ip{\boldsymbol{l}}{\boldsymbol{U}}\vert ^2$  in green, and their sum, in blue, for $\stat{\boldsymbol{U}}$ being one of the eight states  $\stat{\boldsymbol{A}},\cdots,\stat{\boldsymbol{H}}$ respectively. 
It is evident that in all the six walk-states, the sum of the field intensity components, i.e., the blue plot, remains at $1$ at all azimuthal angles, whereas in case of the non-walk-states $\stat{\boldsymbol{E}}$ and $\stat{\boldsymbol{F}}$, the total field intensity varies with the azimuthal angle, exceeding $1$ at some azimuth and subceeding $1$ at other azimuths.
In the rest of this section, we shall focus our attention to such normalized vector beams only.
The aim now is to generate vector beams using only waveplates and q-plates. 
With respect to DTQWs, the standard q-plates mimic the standard shift operator, Eq. (\ref{eq:standard_Shift_Operator}), except that here apart from altering the OAM of the light, the standard q-plate also swaps the circular polarization components. A q-plate with $\Gamma\neq \pi$ functions like a standard q-plate but only on a fraction $\sin\frac{\Gamma}{2}$ of the incident light beam.  Currently, q-plates whose retardance $\Gamma$ can be tuned by varying the applied voltage have been designed \cite{piccirillo2010photon,d2015arbitrary} and are also commercially available \cite{arcoptix}. 
In this context, the effective behavior of a collection of three waveplates and q-plates has been explored in Ref. \cite{delaney2017arithmetic} and Ref. \cite{kadiri2019wavelength}.
To make connection between the q-plate and the quantum step of Eq. (\ref{eq:Step_definition}), we have the following relation: $\hat{q}_\Gamma\left(q,\alpha_0\right)=\hat{T}_\Gamma\left(2\alpha_0,2q,\boldsymbol{l},\boldsymbol{l}\right)$.
It is possible to implement the quantum step $\hat{T}_\pi$, that $T_\Gamma$ of Eq. (\ref{eq:Step_definition}) with $\Gamma=\pi$, using a pair of waveplates and a single q-plate:
\begin{equation}
	\textstyle{\hat{T}_\pi(\delta,p,\boldsymbol{s},\boldsymbol{c})}\equiv \hat{p}_{_{\Gamma_2}}(\alpha_2)\hat{q}_{_\pi}(\frac{p}{2},\alpha_0)\hat{p}_{_{\Gamma_1}}(\alpha_1)
	\label{eq:three_plate_equivalence}
\end{equation}
where the parameters of the plate are given as:
\begin{equation}
	\begin{aligned}
		\Gamma_1&=2\tan^{-1}\left(\frac{\vert \ip{\boldsymbol{r}}{\boldsymbol{c}}\vert}{\vert \ip{\boldsymbol{l}}{\boldsymbol{c}}\vert}\right),\\
		\alpha_1&=\frac{\pi}{2}+\frac{1}{2}\left(\Phi(\ip{\boldsymbol{r}}{\boldsymbol{c}})-\Phi(\ip{\boldsymbol{l}}{\boldsymbol{c}})\right),\\
		\alpha_0&=\frac{1}{2}(\delta-\Phi(\ip{\boldsymbol{l}}	{\boldsymbol{c}})-\Phi(\ip{\boldsymbol{l}}	{\boldsymbol{s}})),\\
		\Gamma_2&=2\tan^{-1}\left(\frac{\vert \ip{\boldsymbol{r}}{\boldsymbol{s}}\vert}{\vert \ip{\boldsymbol{l}}{\boldsymbol{s}}\vert}\right),\\
		\alpha_2&=\frac{1}{2}\left(\Phi(\ip{\boldsymbol{r}}{\boldsymbol{s}})-\Phi(\ip{\boldsymbol{l}}{\boldsymbol{s}})\right)\\
	\end{aligned}
	\label{eq:three_plate_values}
\end{equation}
The three-plate gadget involving a $q_\pi$ plate introduced through Eqs. (\ref{eq:three_plate_equivalence}) and (\ref{eq:three_plate_values}) can reproduce any $\hat{T}_\pi$ quantum step. Replacing the $q_\pi$ plate with a $q_\Gamma$ plate, it can even reproduce any $\hat{T}_\Gamma$ quantum step of Eq. (\ref{eq:Step_definition}), but only if $\Phi(\ip{\boldsymbol{l}}{\boldsymbol{c}})=\Phi(\ip{\boldsymbol{l}}{\boldsymbol{s}})$.
This may appear as an obstacle in our attempt to  obtain desired vector beams by means of quantum walks. 
However, recall that the shrinking algorithm places no restriction on the shift state $\stat{\boldsymbol{s}}$. We can exploit this freedom on $\stat{\boldsymbol{s}}$, not only to overcome the restriction on phase, but also to get rid of one of the plates altogether.  The choice of $\stat{\boldsymbol{s}}=e^{i\Phi(\ip{\boldsymbol{l}}	{\boldsymbol{c}})}\stat{\boldsymbol{l}}$ satisfies the phase relation and also ensures that $\Gamma_2=0$, so that it need not be used at all, and one can work with only a pair of homogeneous waveplate and $q=\frac{1}{2}$ q-plate per step.
A gadget for varying Berek Plate \cite{born2013principles} \begin{figure*}[htbp]
	\centering
	\includegraphics[height=\textheight,width=\linewidth,keepaspectratio]{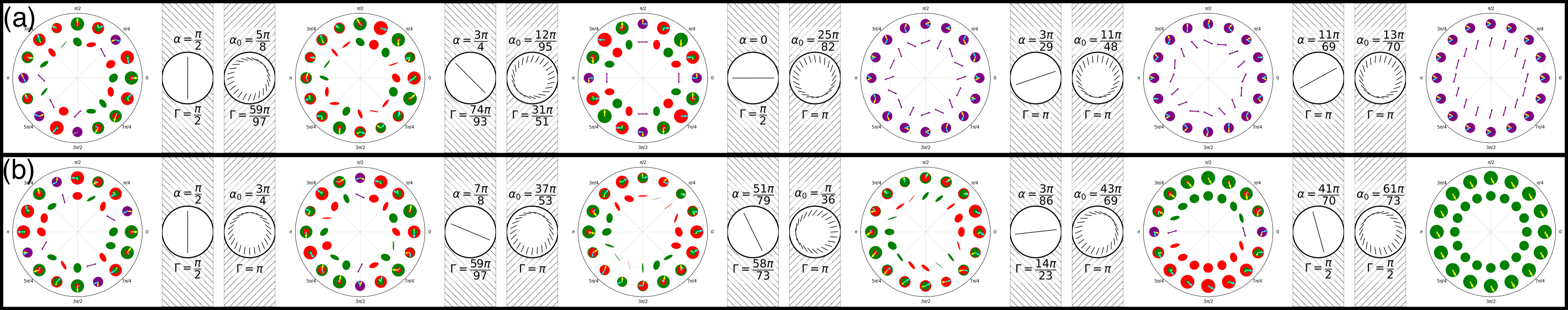}
	\caption{Conversion of the vector beams $\stat{\boldsymbol{G}}$ and $\stat{\boldsymbol{H}}$ of Eqs (\ref{eq:state_G}) and (\ref{eq:state_H}) to standard scalar beams, using five waveplate and q-plate pairs. The first row depicts the progression of the state $\stat{\boldsymbol{G}}$ and the second row depicts that of $\stat{\boldsymbol{H}}$. }
	\label{fig:OAM_G_H_Reverse_Steering}
\end{figure*}

\subsection{Illustration of targeted quantum walks}
We first provide an illustration of the shrinking algorithm for transforming the states $\stat{\boldsymbol{G}}$ and $\stat{\boldsymbol{H}}$ of Eqs. (\ref{eq:state_G}) and (\ref{eq:state_H}) into standard scalar light beams. Recall that $N=\textbf{max}(\vert b \vert,\vert e \vert)$ of both the states is $5$, hence they both can be reduced in five quantum steps, that is, five pair of waveplates and q-plates. The characteristic vectors of these states  are  $\overrightarrow{p}(\boldsymbol{G})=\left(0,\frac{1}{5},\frac{1}{5},\frac{1}{5},\frac{1}{5},\frac{1}{5}\right)$ and  $\overrightarrow{p}(\boldsymbol{H})=\left(\frac{1}{10},\frac{1}{5},\frac{1}{5},\frac{1}{5},\frac{1}{5},\frac{1}{10}\right)$ respectively, and therefore they cannot be output of the same walk for different home-states. 
 Fig. (\ref{fig:OAM_G_H_Reverse_Steering}) shows the transverse plane polarization profiles of both the beams, as they progress through each pair of plates. The $\Gamma$ and $\alpha/\alpha_0$ of these plates are so chosen that every pair of plates shrink the OAM span of the vector beam by one unit from either end.\\
\begin{figure*}[htbp]
	\centering
	\includegraphics[height=\textheight,width=\linewidth,keepaspectratio]{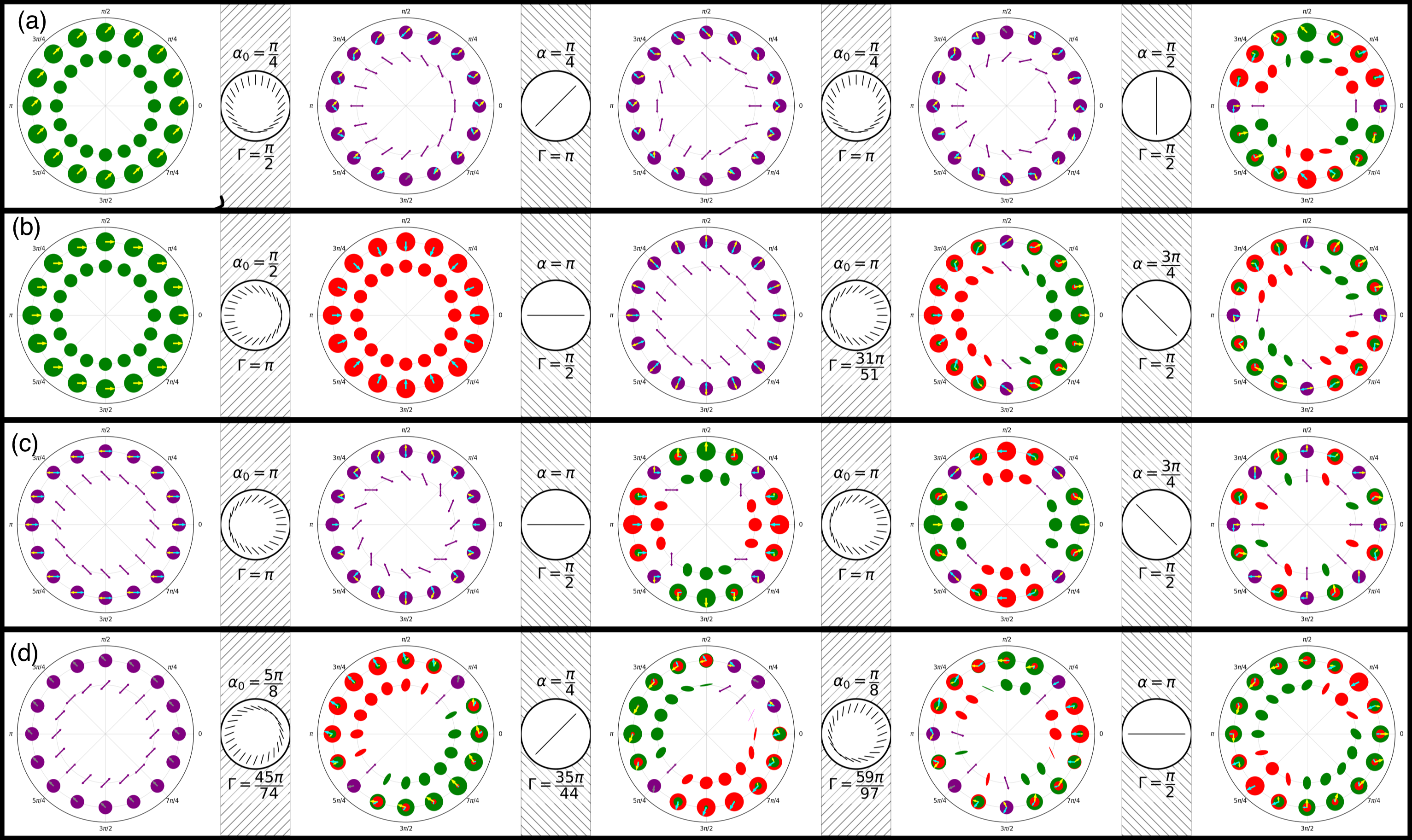}
	\caption{Evolution of scalar beams,  into each of the four vector beams $\stat{\boldsymbol{A}}$, $\stat{\boldsymbol{B}}$, $\stat{\boldsymbol{C}}$, and $\stat{\boldsymbol{D}}$ of Eq \ref{eq:six_states}. }
	\label{fig:OAM_Four_States}
\end{figure*}
We now look to generate the vector beams corresponding to the four walk states $\stat{\boldsymbol{A}},\cdots,\stat{\boldsymbol{D}}$ of Eq. (\ref{eq:six_states}). 
Since $N=2$ for all the four states, each of them can be generated from standard scalar light beams by using just two pairs of waveplates and q-plates.  This conversion is demonstrated in Fig. (\ref{fig:OAM_Four_States}).
\clearpage
\section{Conclusion \label{sec:Conclusion}}
In this work we have considered a bipartite quantum system comprising of a qudit degree of freedom, coupled with a qubit degree of freedom. In such a composite system, we have examined the set of quantum states that are accessible by quantum walks starting from some easy-to-prepare product state. 
A major result coming out of the current work is that such broadest generalization of the definition of quantum walks does not give access to all the states on the composite Hilbert space. A simple criterion for identifying whether a given quantum state can be accessible through a quantum walk is provided based on the notion of translational invariance of a quantum state. Given such a walk-accessible state, a clear and deterministic algorithm for realizing it through discrete time quantum walks is also provided. The given construction is ``minimal'', in that the given composite state cannot be obtained using fewer quantum steps of unit step size than what is given by the algorithm. To give a practical context to our theoretical results, in this paper, we have interpreted them in the context of quantum walks on the Spin-and-Orbital angular momenta of light beams. We have  demonstrated that, any azimuthally varying vector beam can be generated deterministically in such a quantum walk comprising of alternating waveplates and $q=\frac{1}{2}$ q-plates. We believe that the results presented here will aid in designing better quantum-walk based algorithms for quantum state engineering, particularly in the linear optical implementation of discrete-time quantum walks.

\appendix
\section{$\Gamma$ and $\alpha$ for the shrinking algorithm}
Here, given a walk-state $\stat{\boldsymbol{U}}$ of position span $[b,e]$, we find the paramters of the quantum step $\hat{T}_\Gamma(\delta,p,\boldsymbol{s},\boldsymbol{c})$ such that the composite state $\hat{T}_\Gamma(\delta,p,\boldsymbol{s},\boldsymbol{c})\stat{\boldsymbol{U}}$ has a position span of $[b+1,e-1]$.
Let $\stat{\boldsymbol{R}}=\hat{T}_\Gamma(\delta,p,\boldsymbol{s},\boldsymbol{c})\stat{\boldsymbol{U}}$. The amplitude at the $m^{th}$ position $r_m$ and the coin-state there $\stat{\boldsymbol{r_m}}$ can be obtained by Eq. (\ref{eq:Step_definition}) as 
\begin{widetext}
	\begin{equation}
	\begin{aligned}
	r_m\stat{\boldsymbol{r}_m}=&\left(u_m\ip{\boldsymbol{c}}{\boldsymbol{u}_m}\cos\frac{\Gamma}{2}-u_{m+1}\ip{\boldsymbol{c}_\perp}{\boldsymbol{u}_{m+1}}\sin\frac{\Gamma}{2}e^{-i\delta}\right)\stat{\boldsymbol{s}}
	+\left(u_m\ip{\boldsymbol{c}_\perp}{\boldsymbol{u}_m}\cos\frac{\Gamma}{2}+u_{m-1}\ip{\boldsymbol{c}}{\boldsymbol{u}_{m-1}}\sin\frac{\Gamma}{2}e^{i\delta}\right)\stat{\boldsymbol{s}_\perp}
	\end{aligned}
	\end{equation}
\end{widetext}
We chose the parameters of $\hat{T}_\Gamma(\delta,1,\boldsymbol{s},\boldsymbol{c})$ such that (i) $(e+1)^{th}$ and $e^{th}$ positions of $\stat{\boldsymbol{R}}$ are unoccupied, that is, $r_{e+1}=0$ and $r_{e}=0$, and (ii) the $b^{th}$ and ${(b-1)}^{th}$ positions of $\stat{\boldsymbol{R}}$ are also unoccupied, that is, $r_{b-1}=0$ and $r_{b}=0$. 
\begin{equation}
\begin{aligned}
\ip{\boldsymbol{c}}{\boldsymbol{u_e}}&=0,\\
\tan\frac{\Gamma}{2}&=\frac{u_e\vert\ip{\boldsymbol{c}_\perp}{\boldsymbol{u}_e}\vert}{u_{e-1}\vert\ip{\boldsymbol{c}}{\boldsymbol{u}_{e-1}}\vert},\\
\delta&=\pi+\Phi(\ip{\boldsymbol{c}_\perp}{\boldsymbol{u}_e})-\Phi(\ip{\boldsymbol{c}}{\boldsymbol{u}_{e-1}})
\label{eq:right_shrink}
\end{aligned}
\end{equation}Note that we are dealing with quantum walk steps of unit size, $p=1$. The contribution to the amplitude at the $(e+1)^{th}$ position of $\stat{\boldsymbol{R}}$ is therefore only from that of $e^{th}$ position of $\stat{\boldsymbol{U}}$ alone. The first condition ensures that this contribution is $0$. The amplitude of $e^{th}$ component of $\stat{\boldsymbol{R}}$, on the other hand, gets contributions from both $e^{th}$ and $(e-1)^{th}$ components of $\stat{\boldsymbol{U}}$. The chosen $\Gamma$ and $\alpha$ of $\hat{T}$ are such that the two contributions are equal and out of phase, so that they cancel each other. \\
In a similar manner, the condition that the $b^{th}$ and $(b-1)^{th}$ positions are unoccupied in $\stat{\boldsymbol{R}}$ leads to the following three conditions on the step parameters:
\begin{equation}
\begin{aligned}
\ip{\boldsymbol{c}_\perp}{\boldsymbol{u_b}}&=0,\\
\tan\frac{\Gamma}{2}&=\frac{u_b\vert\ip{\boldsymbol{c}}{\boldsymbol{u}_b}\vert}{u_{b+1}\vert\ip{\boldsymbol{(u_b)}_\perp}{\boldsymbol{u}_{b+1}}\vert},\\
\delta&=\Phi(\ip{\boldsymbol{c}}{\boldsymbol{u}_b})-\Phi(\ip{(\boldsymbol{u_b})_\perp}{\boldsymbol{u}_{b+1}})
\end{aligned}
\label{eq:left_shrink}
\end{equation}
The contribution to $(b-1)^{th}$ of $\stat{\boldsymbol{R}}$ is from $b^{th}$ component of $\stat{\boldsymbol{U}}$ alone, and the first condition ensures that this contribution is $0$. The $b^{th}$ component of $\stat{\boldsymbol{R}}$, gets contributions from both the $b^{th}$ and $(b-1)^{th}$ components of $\stat{\boldsymbol{U}}$ and the $\Gamma$ and $\alpha$ of the step $\hat{T}$ ensure that these two contributions cancel each other. 
The step of Eq. (\ref{eq:right_shrink}), acting on a composite state $\stat{\boldsymbol{U}}$ changes its position spread from $[b,e]$ to $[b-1,e-1]$ and, likewise, the step of Eq. (\ref{eq:left_shrink}) changes the position span from $[b,e]$ to $[b+1,e+1]$. We are interested in shrinking the span to $[b+1,e-1]$,  so we seek a step whose parameters satisfy both sets of constraints simultaneously. For instance, the first of Eq. (\ref{eq:right_shrink})  and the first of Eq. (\ref{eq:left_shrink}) can be simultaneously satisfied only if $\ip{\boldsymbol{u}_b}{\boldsymbol{u}_e}=0$. This is not guaranteed for any arbitrary composite state of the form Eq. (\ref{eq:arb_composite_state}). We, however, are not dealing with arbitrary composite states, but the ``walk-states" satisfying Eq. (\ref{eq:TI_Constraints}), and for such states, it can be shown that the two set of conditions are indeed identical. We make a choice of $\stat{\boldsymbol{c}}=\stat{(\boldsymbol{u}_e)_\perp}$, from which Eq. (\ref{eq:shrinking_plate_parameters}) follows. 

\clearpage

	
	\bibliographystyle{apsrev4-1}
	\bibliography{Q_Walk}

\end{document}